\documentclass[12pt]{article}
\usepackage{amssymb,amsmath,epsfig}

\begin{document}

\title{\bf Cosmological Analysis of Reconstructed $\mathcal{F}(T,T_\mathcal{G})$ Models}
\author{M. Sharif \thanks{msharif.math@pu.edu.pk} and
Kanwal Nazir \thanks{awankanwal@yahoo.com}~~\thanks{On leave from
Department of Mathematics, Lahore College
for Women University, Lahore-54000, Pakistan.}\\
Department of Mathematics, University of the Punjab,\\
Quaid-e-Azam Campus, Lahore-54590, Pakistan.}

\date{}
\maketitle
\begin{abstract}
In this paper, we analyze cosmological consequences of the
reconstructed generalized ghost pilgrim dark energy
$\mathcal{F}(T,T_\mathcal{G})$ models in terms of redshift parameter
$z$. For this purpose, we consider power-law scale factor, scale
factor for two unified phases and intermediate scale factor. We
discuss graphical behavior of the reconstructed models and examine
their stability analysis. Also, we explore the behavior of equation
of state as well as deceleration parameters and
$\omega_{\Lambda}-\omega_{\Lambda}^{'}$ as well as $r-s$ planes. It
is found that all models are stable for pilgrim dark energy
parameter $2$. The equation of state parameter satisfies the
necessary condition for pilgrim dark energy phenomenon for all scale
factors. All other cosmological parameters show great consistency
with the current behavior of the universe.
\end{abstract}

\section{Introduction}

The accelerated cosmic expansion phenomenon is undoubtedly the
biggest achievement of the twentieth century. The source behind this
expansion is said to be a repulsive force called dark energy (DE)
having large negative pressure. This energy is evenly scattered in
the universe but we do not know much about its nature as well as
composition. It is suggested by WMAP experiment that the universe
has budget as $73\%$ DE, $23\%$ dark matter and $4\%$ baryonic
matter. The cosmic expansion goes through different stages of dark
matter and DE normally characterized by the equation of state (EoS)
parameter. These ranges include $\omega<-1$ for phantom, $\omega=-1$
for vacuum (cosmological constant $(\Lambda)$) and
$-1<\omega<\frac{1}{3}$ for quintessence DE dominated eras. The
matter dominated eras corresponding to $\omega=0, 1$ and
$\frac{1}{3}$ represent cold dust, stiff and radiation dominated
eras, respectively.

The cosmological constant is the best ingredient to discuss the DE
mystery but it has issues like coincidence and fine-tuning. This
motivates researchers to find some alternatives to describe the DE
nature. The most appealing proposals in this scenario are either to
modify matter or gravitational side of the Einstein-Hilbert action.
The matter modification provides different DE models like Chaplygin
gas, phantom, quintessence, k-essence, holographic and pilgrim dark
energy (PDE) etc \cite{1}-\cite{1d}. On the other hand, the
gravitational modification leads to modified theories. Among these
theories, there is a modification based on torsional formation of
general relativity dubbed as teleparallel theory. In this theory,
the basic entity is torsion instead of curvature.

The generalization of teleparallel theory is known as
$\textit{f(T)}$ theory in which torsion scalar $T$ is replaced by an
arbitrary function $\textit{f(T)}$ in the action. Recently, an
extension of this theory is proposed by introducing teleparallel
equivalent Gauss-Bonnet term $T_\mathcal{G}$ known as
$\mathcal{F}(T,T_\mathcal{G})$ theory. The reason behind this
generalization is to develop an action that includes higher torsion
correction terms. Kofinas and Saridakis \cite{6} presented this
distinct theory by evaluating a torsion equivalent of Gauss-Bonnet
term without using curvature formalism. They analyzed several
observable like EoS, DE density and matter density parameters by
assuming two specific $\mathcal{F}(T,T_\mathcal{G})$ models
\cite{6a} and found this theory as explaining the cosmic evolution.

Kofinas et al. \cite{6b} discussed dynamical analysis of spatially
flat FRW metric by considering a particular
$\mathcal{F}(T,T_\mathcal{G})$ model and concluded that the universe
can exhibit various DE dominated solutions such as cosmological
constant, quintessence or phantom like solutions that depend upon
the values of corresponding model parameters. Waheed and Zubair
\cite{11a} investigated energy bounds with perfect fluid using
Hubble, deceleration, jerk and snap parameters. Zubair and Jawad
\cite{6c} explored laws of thermodynamics at apparent horizon of FRW
metric. Jawad \cite{11b} studied energy conditions for FRW universe
analytically.

Various DE models have been developed in the context of quantum
gravity as well as general relativity. One of the them is the
Veneziano ghost DE model defined as $\rho_{T}=\mu H$, where $\mu$ is
a constant \cite{11c}. This model is interesting as it does not
involve any new degree of freedom or new parameter. The Veneziano
ghost energy density has the form $H+O(H^2)$ that provides enough
amount of vacuum energy to explore the expansion phenomenon.
However, ghost DE model involves only the term $H$ in its energy
density. Therefore, Cai et al. \cite{13} added the term $H^2$ in the
ghost DE model as $\rho_{T}=\mu H+\nu H^{2}$, where $\nu$ is another
constant, known as generalized ghost DE model. Fernandez \cite{5b}
discussed ghost DE models along with scalar field whereas Malekjani
\cite{5c} established different $\textit{f(R)}$ models considering
ghost as well as generalized ghost DE models.

Wei \cite{12} presented another DE model dubbed as pilgrim DE (PDE)
motivated by the fact that phantom like DE has enough strength to
prevent black hole formation rather than other types of DE. The PDE
also encouraged this fact due to its same repulsive nature. The
generalized ghost DE density is further established by involving PDE
parameter as
\begin{equation}\label{1}
\rho_{T}=(\mu H+\nu H^{2})^{u},
\end{equation}
where $u$ represents PDE parameter. The generalized ghost DE model
after involving PDE parameter is named as generalized ghost PDE
(GGPDE) model. Sharif and Jawad \cite{5d} examined flat FRW model
for interacting as well as non-interacting GGPDE model and found
that this model fulfills PDE phenomenon.

The reconstruction technique is the most suitable approach to
develop an appropriate DE model which successfully draws the picture
of cosmic history. According to this technique, one has to equate
energy densities of corresponding DE model and modified theory to
derive a reconstructed model. Jawad and Rani \cite{8} discussed
GGPDE model by applying this technique in Horava-Lifshitz $f(R)$
gravity. Sharif and Nazir \cite{8a} worked on this technique by
assuming GGPDE $f(T)$ model and investigated the behavior of
different cosmological parameters. Jawad et al. \cite{10}
investigated ghost DE model in $\mathcal{F}(T,T_\mathcal{G})$
gravity and examined its cosmological consequences through the
reconstructed model. Sharif and Nazir \cite{10a} reconstructed GGPDE
$\mathcal{F}(T,T_\mathcal{G})$ models and discussed their
corresponding EoS parameters versus PDE parameter.

In this paper, we study cosmological behavior of the reconstructed
models \cite{10a} versus redshift parameter $z$ and discuss their
stability through squared speed of sound parameter. We investigate
these reconstructed models through EoS parameter, deceleration
parameter, $\omega_\Lambda-\omega_\Lambda'$ analysis and $r-s$
plane. The paper is arranged as follows. Next section provides basic
introduction of $\mathcal{F}(T,T_\mathcal{G})$ gravity. In section
\textbf{3}, we briefly describe the well-known scale factors.
Section \textbf{4} analyzes the evolution trajectories via
cosmological parameters. In the last section, we summarize the
results.

\section{$\mathcal{F}(T,T_\mathcal{G})$ Gravity}

In this section, we provide a concise review of
$\mathcal{F}(T,T_\mathcal{G})$ gravity in the background of FRW
geometry. The tetrad field $e_{A}(x^{\alpha})$ has a fundamental
role in $f(T)$ as well as $\mathcal{F}(T,T_\mathcal{G})$ gravity.
Trivial tetrad is the simplest one expressed as
$e_{A}=\partial_{\alpha}{\delta^{\alpha}}_{A}$ and
$e^{B}=\partial^{\alpha}{\delta_{\alpha}}^{B}$, where
$\delta^{\alpha}_{a}$ is the Kronecker delta. These are not commonly
used because they provide zero torsion. The non-trivial tetrad have
different behavior, so they are more supportive in describing
teleparallel theory. These tetrad can be represented as
\begin{equation}\nonumber
h_{A}=\partial_{\alpha}{h_{A}^{~\alpha}},\quad
h^{B}=dx^{\alpha}{h^{B}_{~\alpha}},
\end{equation}
satisfying
\begin{equation}\nonumber
h_{~\alpha}^{A}h^{~\alpha}_{B}=\delta^{A}_{B},\quad
h_{~\alpha}^{A}h^{~\beta}_{A}=\delta^{\beta}_{\alpha}.
\end{equation}
The metric tensor can also be expressed in the product of tetrad
fields as
\begin{equation}\nonumber
g_{\alpha\beta}=\eta_{AB}h_{\alpha}^{A}h^{B}_{\beta},
\end{equation}
where $\eta_{AB}$~=~diag$(1,-1,-1,-1)$ is the Minkowski metric. The
coordinates on manifold are represented by Greek indices
$(\alpha,\beta,...)$ while coordinates on tangent space are
characterized by Latin indices $(A, B,...)$.

The Weitzenb$\ddot{o}$ck connection ${\omega^{A}}_{B}(x^{\alpha})$
that describes parallel transportation, has the following form
\begin{equation}\nonumber
\omega^{\beta}_{\alpha\gamma}={h^{\beta}}_{A}{h^{A}}_{\alpha,\gamma}.
\end{equation}
The structure coefficients $\mathcal{C}^{C}_{AB}$ are defined as
\begin{equation}\nonumber
[h_{A},h_{B}]=h_{C}\mathcal{C}^{C}_{AB},
\end{equation}
where
\begin{equation}\nonumber
\mathcal{C}^{C}_{AB}={h^{\beta}}_{B}{h^{\alpha}}_{A}
({h^{C}}_{\alpha,\beta}-{h^{C}}_{\beta,\alpha}).
\end{equation}
Similarly, we can express the torsion as well as curvature tensors
as
\begin{eqnarray}\nonumber
T^{A}_{BC}&=&-\omega^{A}_{BC}+\omega^{A}_{CB}-\mathcal{C}^{A}_{BC},\\\nonumber
R^{A}_{BCD}&=&-\omega^{E}_{BC}
\omega^{A}_{ED}+\omega^{A}_{BD,C}+\omega^{E}_{BD}\omega^{A}
_{EC}-\mathcal{C}^{E}_{CD}\omega^{A}_{BE}-\omega^{A}_{BC,D}.
\end{eqnarray}
The contorsion tensor is defined by
\begin{equation}\nonumber
\mathcal{K}_{ABC}=\frac{1}{2}(-T_{BCA}-T_{ABC}+T_{CAB})=-\mathcal{K}_{BAC}.
\end{equation}
Finally, the torsion scalars $T$ and $T_\mathcal{G}$ take the form
\begin{eqnarray}\nonumber
T&=&\frac{1}{4}T^{ABC}T_{ABC}-T_{AB}^{~~A}T^{CB}_{~~C}
+\frac{1}{2}T^{ABC}T_{CBA},\\\nonumber
T_\mathcal{G}&=&(2{{\mathcal{K}^{A_{3}}}_{EB}\mathcal{K}^{A_{1}A_{2}}}_{A}
{\mathcal{K}^{EA_{4}}}_{F}{\mathcal{K}^{F}}_{CD}
+{\mathcal{K}^{A_{2}}}_{B}{\mathcal{K}^{A_{1}}}_{EA}
{\mathcal{K}^{A_{3}}}_{FC}{\mathcal{K}^{FA_{4}}}_{D}
+2{\mathcal{K}^{A_{3}}}_{EB}
\\\nonumber&\times&{\mathcal{K}^{A_{1}A_{2}}}_{A}
{\mathcal{K}^{EA_{4}}}_{C,D}-2{{\mathcal{K}^{A_{3}}}_{EB}
\mathcal{K}^{A_{1}A_{2}}}_{A}
{\mathcal{K}^{E}}_{FC}{\mathcal{K}^{FA_{4}}}_{D})
\delta^{ABCD}_{A_{1}A_{2}A_{3}A_{4}}.
\end{eqnarray}

The action for $\mathcal{F}(T,T_\mathcal{G})$ gravity is proposed by
Kofinas and Saridakis \cite{6}
\begin{equation}\nonumber
S=\int
d^4xh\left[\frac{\mathcal{F}(T,T_\mathcal{G})}{2\kappa^{2}}
+\mathcal{L}_{m}\right],\quad
h=\det(h^{I}_\beta)=\sqrt{|g|},
\end{equation}
where $\mathcal{L}_{m}$ is the matter Lagrangian and $\kappa^{2}=1$.
The teleparallel equivalent to general relativity is obtained by
substituting $\mathcal{F}(T,T_\mathcal{G})=-T$. We can also have
Gauss-Bonnet theory when $\mathcal{F}(T,T_\mathcal{G})=\alpha
T_{\mathcal{G}}-T$, where $\alpha$ represents Gauss-Bonnet coupling.
The $\mathcal{F}(T,T_\mathcal{G})$ field equations can be obtained
by varying the action as
\begin{eqnarray}\nonumber
&&2(H^{[AC]B}-H^{[CB]A}+H^{[BA]C})_{,C}+2(H^{[AC]B}
-H^{[CB]A}+H^{[BA]C})C_{~DC}^{D}
\\\nonumber&+&(2H^{[AC]D}+H^{DCA})C^{B}_{~CD}+4H^{[DB]C}C_{(DC)}^{~~~A}
+T^{A}_{~CD}H^{CDB}-\mathcal{H}^{AB}+(F\\\label{C}&
-&TF_{T}-T_{G}F_{T_{G}})\eta^{AB}=\mathcal{T}^{AB},
\end{eqnarray}
where
\begin{eqnarray}\nonumber
H^{ABC}&=&F_{T}(\eta^{AC}\mathcal{K}^{BD}_{~~D}-\mathcal{K}^{BCA})
+F_{T_{G}}[\epsilon^{CPRI}
(2\epsilon^{A}_{~DKE}\mathcal{K}^{BK}_{~~P}\mathcal{K}^{D}_{~QR}
+\epsilon_{QDKE}\\\nonumber&\times&\mathcal{K}^{AK}_{~~P}
\mathcal{K}^{BD}_{~~R}
+\epsilon^{AB}_{~~KE}\mathcal{K}^{K}_{~DP}\mathcal{K}^{D}_{~QR})
\mathcal{K}^{QE}_{~~I}
+\epsilon^{CPRI}\epsilon^{AB}_{~~KD}\mathcal{K}^{ED}_{~~P}
(\mathcal{K}^{K}_{~ER,I}
\\\nonumber&-&\frac{1}{2}\mathcal{K}^{K}_{~EQ}C^{Q}_{~IR})
+\epsilon^{CPRI}\epsilon^{AK}_{~~DE}\mathcal{K}^{DE}_{~P}
(\mathcal{K}^{B}_{~KR,I}
-\frac{1}{2}\mathcal{K}^{B}_{~KQ}C^{Q}_{~IR})]
+\epsilon^{CPRI}\\\nonumber&\times&\epsilon^{A}_{~KDE}
[(F_{T_{G}}\mathcal{K}^{BK}_{~P}\mathcal{K}^{DE}_{~R})_{,I}
+F_{T_{G}}C^{Q}_{~PI}
\mathcal{K}^{BK}_{~~[Q}\mathcal{K}^{DE}_{~~R]}],
\end{eqnarray}
and
\begin{equation}\nonumber
\mathcal{H}^{AB}=F_{T}\epsilon^{A}_{~KCE}\epsilon^{BRIE}K^{K}_{~ER}K^{EC}_{~~I},
\end{equation}
$\epsilon^{1234}=-1$,~ $\epsilon_{1234}=1$ and $0$ otherwise. Also,
$F_{T}=\frac{\partial F}{\partial T}$, $F_{T_{G}}=\frac{\partial
F}{\partial T_{G}}$ and $\mathcal{T}^{AB}$ is the energy-momentum
tensor.

Now we discuss cosmological significance of
$\mathcal{F}(T,T_\mathcal{G})$ theory by considering flat FRW
universe model as
\begin{equation}\nonumber
ds^2=-dt^2+a^2(t)(dx^2+dy^2+dz^2),
\end{equation}
where $a(t)$ is the scale factor. There exist infinite possible
tetrad fields for each metric, thus we choose a common tetrad field
for FRW metric as
\begin{equation}\label{2}
h^{I}_{~\alpha}=diag(1,a(t),a(t),a(t)),
\end{equation}
where dual is defined as
\begin{equation}\\\nonumber
h_{I}^{~\alpha}=diag(1,a^{-1}(t),a^{-1}(t),a^{-1}(t)).
\end{equation}
The corresponding torsion scalars are
\begin{equation}\label{4}
T=6H^{2},~~ T_{G}=24H^{2}(\dot{H}+H^{2}).
\end{equation}
Here, $H=\frac{\dot{a}}{a}$ defines Hubble parameter and dot
indicates time derivative. Substituting the above values in
Eq.(\ref{C}), we obtain
\begin{eqnarray}\label{5}
\rho&=&\frac{1}{2}\left[\mathcal{F}+6H^2-T_{G}\mathcal{F}_{T_{G}}-12H^{2}\mathcal{F}_{T}
+24H^{3}\dot{\mathcal{F}}_{T_{G}}\right],\\\nonumber
-p&=&\frac{1}{2}\left[\mathcal{F}+2(2\dot{H}+3H^2)-4(3H^{2}+\dot{H})
\mathcal{F}_{T}-4H\dot{\mathcal{F}}_{T}-T_{G}\mathcal{F}_{T_{G}}
\right.\\\label{6}&+&\left.\frac{2}{3H}\dot{\mathcal{F}}_{T_{G}}T_{G}
+8\ddot{\mathcal{F}}_{T_{G}}H^{2}\right].
\end{eqnarray}
We can rewrite the above equations in usual form as
\begin{eqnarray}\nonumber
3H^{2}&=&\rho_{m}+\rho_{\Lambda},\\\nonumber
2\dot{H}&=&-(\rho_{m}+\rho_{\Lambda}+p_{m}+p_{\Lambda}),
\end{eqnarray}
where the energy density and pressure for DE sector are
\begin{eqnarray}\label{10}
\rho_{\Lambda}&=&-\frac{1}{2}(\mathcal{F}-T_{G}\mathcal{F}_{T_{G}}
-12H^{2}\mathcal{F}_{T}+24H^{3}\dot{\mathcal{F}}_{T_{G}}),\\\nonumber
p_{\Lambda}&=&\frac{1}{2}(\mathcal{F}
-4H\dot{\mathcal{F}}_{T}-4(\dot{H}+3H^{2})\mathcal{F}_{T}-T_{G}\mathcal{F}_{T_{G}}
+8H^{2}\ddot{\mathcal{F}}_{T_{G}}\\\label{11}
&+&\frac{2}{3H}T_{G}\dot{\mathcal{F}}_{T_{G}}).
\end{eqnarray}
The energy conservation equations in terms of dark matter and DE are
\begin{eqnarray}\nonumber
\dot{\rho}_{m}+3H(p_{m}+\rho_{m})&=&0,\\\nonumber
\dot{\rho}_{\Lambda}+3H(p_{\Lambda}+\rho_{\Lambda})&=&0.
\end{eqnarray}

\section{Cosmic Scale Factors}

Here, we briefly describe some scale factors through which we
explore the cosmological behavior of our reconstructed models
\cite{10a}.

\begin{itemize}
\item \textbf{Power-Law Scale Factor}
\end{itemize}

This scale factor is defined as \cite{23}
\begin{equation}\nonumber
a(t)=a_{0}t^{n},
\end{equation}
where $n>0$, $a_0>0$. This form of scale factor provides a great
consistency for flat FRW metric with the supernova data. For $n>1$,
it gives an accelerating universe. Using this scale factor, we
obtain the corresponding values as
\begin{eqnarray}\label{12a}
H=\frac{n}{t},\quad T=\frac{6n^2}{t^2},\quad
T_{G}=\frac{24(n-1)n^3}{t^4},
\end{eqnarray}
and Eq.(\ref{1}) becomes
\begin{equation}\nonumber
\rho_T=\left[\mu \left(nt^{-1}\right)+\nu
\left(nt^{-1}\right)^{2}\right]^u =\left[\mu^u\left(nt^{-1}\right)^u
+u\mu^{u-1}\nu\left(nt^{-1}\right)^{u+1}\right].
\end{equation}
\begin{itemize}
\item \textbf{Scale Factor for Unified Phases}
\end{itemize}
The following scale factor unifies matter as well as DE dominated
phases. The Hubble parameter takes the form as \cite{23}-\cite{23b}
\begin{eqnarray}\label{12c}
H(t)=\frac{H_{2}}{t}+H_{1},
\end{eqnarray}
which leads to the following form of scale factor as
\begin{eqnarray}\nonumber
a(t)=a_1t^{H_{2}}e^{H_{1}t}.
\end{eqnarray}
When $t$ is very small, we obtain $H(t)\sim\frac{H_{2}}{t}$ which
exhibits the presence of perfect fluid with
$\omega_{\Lambda}=\frac{2}{3}H_{2}^{-1}-1$. Moreover, when $t$ is
very large, $H\rightarrow H_1$ yielding constant Hubble parameter
which leads to de Sitter universe. This type of Hubble parameter
yields a transition from matter to DE dominated phases. The
corresponding values of torsion scalars are
\begin{eqnarray}\nonumber
T=6\left(\frac{H_{2}}{t}+H_{1}\right)^2,\quad
T_G=24\left(\frac{H_2}{t}+H_1\right)^2\left(\left(\frac{H_{2}}{t}
+H_{1}\right)^2-\frac{H_2}{t^2}\right).
\end{eqnarray}
Using Eqs.(\ref{1}) and (\ref{12c}), it follows that
\begin{eqnarray}\nonumber
\rho_T=\left[\mu \left(\frac{H_{2}}{t}+H_{1}\right)+\nu
\left(\frac{H_{2}}{t}+H_{1}\right)^{2}\right]^u.
\end{eqnarray}
\begin{itemize}
\item \textbf{Intermediate Scale Factor}
\end{itemize}
The scale factor and the corresponding Hubble parameter are defined
as \cite{24}
\begin{eqnarray}\label{12d}
a(t)=\exp(bt^{m}),\quad H(t)=bmt^{m-1},
\end{eqnarray}
where $0<m<1$ and $b$ is an arbitrary constant. This scale factor is
much useful in cosmological analysis as it has great consistency
with astrophysical observations. Both torsion scalars for this scale
factor take the form
\begin{eqnarray}\nonumber
T=6b^2m^2t^{(m-1)},\quad
T_G=24b^2m^2t^{2(m-1)}\left[b^2m^2t^{2(m-1)}+\frac{bmt^{(m-1)}(m-1)}{t}\right].
\end{eqnarray}
The corresponding energy density of GGPDE is
\begin{eqnarray}\nonumber
\rho_T=(\mu bmt^{(m-1))})^u+\nu \mu (bmt^{(m-1)})^{(u+1)}u^{(u-1)}.
\end{eqnarray}

\section{Cosmological Analysis Via Well-Known Scale Factors}

In this section, we explore the behavior of the reconstructed models
and investigate their stability. We discuss EoS as well as
deceleration parameters and $\omega_\Lambda-\omega'_\Lambda$ as well
as $r-s$ planes by using the above three scale factors. We consider
the reconstructed $\mathcal{F}(T,T_\mathcal{G})$ models \cite{10a}
obtained by equating the corresponding energy densities of GGPDE
model and $\mathcal{F}(T,T_\mathcal{G})$ gravity. For this purpose,
we equate Eqs.(\ref{1}) and (\ref{10}), i.e.,
$\rho_{\Lambda}=\rho_{T}$ as
\begin{equation}\label{12b}
-\frac{1}{2}(\mathcal{F}-12H^{2}\mathcal{F}_{T}-T_{G}\mathcal{F}_{T_{G}}
+24H^{3}\dot{\mathcal{F}}_{T_{G}})=(\mu H+\nu H^{2})^u.
\end{equation}
We can determine solution of the above equation only for a
particular choice of the scale factor. Thus we consider all the
above three scale factors and analyze their behavior.

\subsection{Power-Law Scale Factor}

For this scale factor, we substitute Eq.(\ref{12a}) in (\ref{12b})
and then solving the resulting equation, we obtain the reconstructed
GGPDE $\mathcal{F}(T,T_\mathcal{G})$ model as
\begin{eqnarray}\nonumber
\tilde{F}(T,T_\mathcal{G})&=&(4(n-1)t^{\frac{-2n}{n-1}}
(3n^2t^{\frac{2}{n-1}}(4+u(7+u)-n(4+5u))(12+u(9\\\nonumber&+&u)-n(9+5u))
\mu-2\mu^u(7n-11)\left(\frac{n}{t}\right)^ut^{\frac{n}{n-1}}
(t^{\frac{n}{n-1}}(-12-u\\\nonumber&\times&(9+u)+n(9+5u))\mu-mt^{\frac{1}{n-1}}
u(-4-u(7+u)+n(4+5u))\\\nonumber&\times&\nu)))/((7n-11)(4+u(7+u)-n(4+5u))
(12+u(9+u)\\\nonumber&-&n(9+5u))\mu)+c_1t^{\frac{1}{2}[7-5n
-\sqrt{\frac{-(n-1)(33+n(-54+25n))}{1-n}}]}\\\label{15}&+&c_2t^{\frac{1}{2}[7-5n
+\sqrt{\frac{-(n-1)(33+n(-54+25n))}{1-n}}]}.
\end{eqnarray}
Here, we denote $\eta=\sqrt{\frac{-(n-1)(33+n(-54+25n))}{1-n}}$ and
the reconstructed model as $\tilde{F}(T,T_\mathcal{G})$. It is
predicted that phantom-like DE (having repulsive nature) is too
strong to avoid the black hole formation. Wei \cite{12} estimated
that total vacuum energy of a system having size $L$ could cross the
limit of same size black hole mass, i.e., $\rho L^3\geq m_{p}^2L$,
which is the first requirement for PDE. The energy density of PDE
model is defined as $\rho=3n^{2}m_{p}^{4-u}L^{-u}$, where $m_{p}$ is
the Planck mass. Thus, we have $l_{p}^{2-u}=L^{2-u}\geq m_{p}^{u-2}$
which implies that $u\leq2$ for $L\geq l_{p}$, where $l_{p}$ defines
the Planck length. Hence, we examine the evolution of reconstructed
GGPDE $\tilde{F}(T,T_\mathcal{G})$ model for two values of PDE
parameter as $u=2$ and $-2$. For this purpose, we take
$a=a_0(1+z)^{-1}$, where $z$ is the redshift parameter. Also, we
consider the values of model parameters as $\mu=1.55$, $\nu=1.91$
\cite{5d,10a}. We take the remaining parameters as $a_0=1$,
$c_1=0.9$ and $c_2=-0.004$. The plots for reconstructed
$\tilde{F}(T,T_\mathcal{G})$ model versus $n$ as well as redshift
parameter $z$ are displayed in Figure \textbf{1}. It is observed
that the left plot of reconstructed $\tilde{F}(T,T_\mathcal{G})$
model represents increasing pattern for $n\geq3$. In the right plot,
the reconstructed $\tilde{F}(T,T_\mathcal{G})$ model exhibits
decreasing behavior initially, then it becomes flat and at the end,
it shows increasing behavior for $n\geq3.2$.
\begin{figure}\center
\epsfig{file=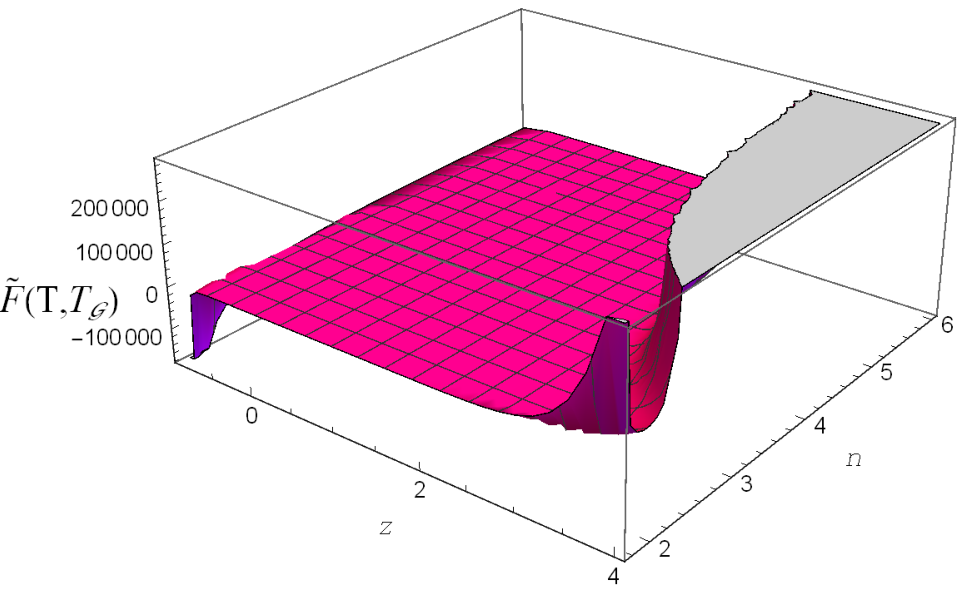,width=0.5\linewidth}
\epsfig{file=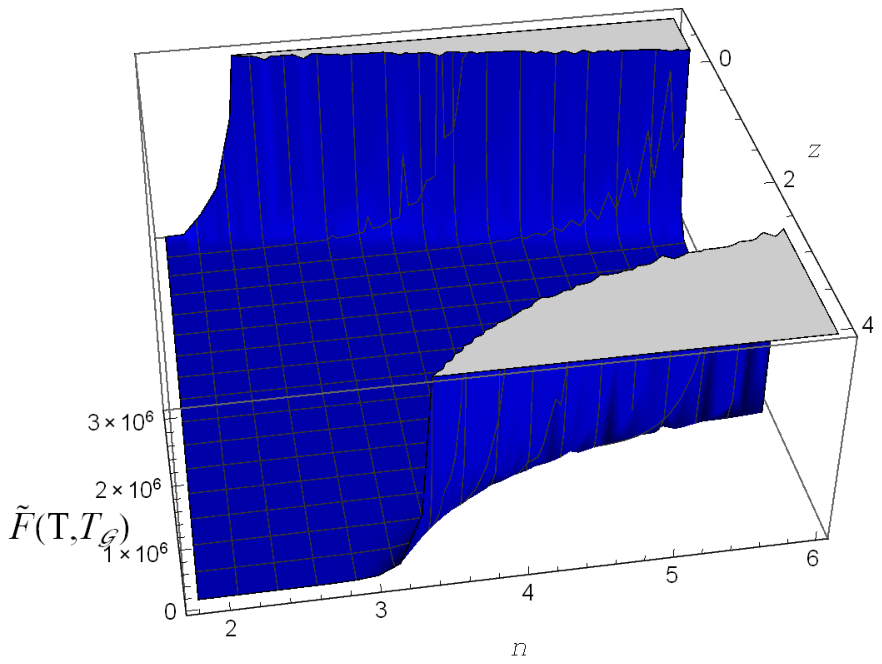,width=0.45\linewidth}\caption{Plots of power-law
reconstructed GGPDE $\tilde{F}(T,T_\mathcal{G})$ model for $u=2$
(left) and $-2$ (right).}
\end{figure}

\begin{itemize}
\item Now, we investigate stability of the reconstructed GGPDE
$\tilde{F}(T,T_\mathcal{G})$ model through the squared speed of
sound $v_s^2$ defined as
\begin{equation}\nonumber
v_s^2=\frac{\dot{p}_\Lambda}{\dot{\rho}_\Lambda}.
\end{equation}
The positive value $(v_s^2>0)$ indicates stability of the model
whereas its negative value $(v_s^2<0)$ corresponds to instability of
the model. Using Eqs.(\ref{10}), (\ref{11}) and (\ref{15}) in the
above expression, we discuss squared speed of sound parameter
graphically. The plots of $v_s^2$ versus $n$ and redshift parameter
$z$ are shown in Figure \textbf{2}. In the left plot of Figure
\textbf{2} $(u=2)$, the squared speed of sound parameter is positive
showing that the reconstructed $\tilde{F}(T,T_\mathcal{G})$ model is
stable. For $u=-2$ (right), the corresponding model is not stable at
present as well as future epoch.
\item The evolutionary behavior of EoS parameter for the reconstructed
$\tilde{F}(T,T_\mathcal{G})$ model is analyzed by evaluating
$\omega_\Lambda$ through Eqs.(\ref{10}) and (\ref{11}) as follows
\begin{eqnarray}\nonumber
\omega_\Lambda&=&-1+(-4\dot{H}-24H^3\dot{\mathcal{F}}_{T_{G}}
-4\dot{H}\mathcal{F}_{T}-4H\dot{\mathcal{F}}_{T}+(2/3)HT_{G}\dot{\mathcal{F}}_{T_{G}}
\\\label{16}&+&8H^{2}\ddot{\mathcal{F}}_{T_{G}})(6H^{2}-\mathcal{F}+12H^{2}\mathcal{F}_{T}
+T_{G}\mathcal{F}_{T_{G}}-24H^{3}\dot{\mathcal{F}}_{T_{G}})^{-1}.
\end{eqnarray}
Substituting Eq.(\ref{15}) in the above expression, we obtain the
EoS parameters for $u=2$ (Figure \textbf{3} left) and $u=-2$ (right)
in terms of $z$. We consider same values of the corresponding
constants as taken earlier. We investigate the evolution of EoS
parameter in the interval $-0.9\leq z\leq2$ for $n_1=3.2$, $n_2=4$
and $n_3=5$. Figure \textbf{3} (left plot) shows that
$\omega_\Lambda$ starts from phantom region, cuts the phantom divide
line and at the end, it becomes zero. This means that EoS parameter
shows quintom behavior for $u=2$. Similarly, the right plot
represents that $\omega_\Lambda$ starts from dust like matter era,
passes via quintessence as well as vacuum DE eras and finally enters
in the phantom era. Hence, $\omega_\Lambda$ behaves like quintom for
$u=-2$. In both cases, the reconstructed
$\tilde{F}(T,T_\mathcal{G})$ model satisfies PDE phenomenon.
\end{itemize}
\begin{itemize}
\item
The deceleration parameter $q$ is described as
\begin{equation}\label{17}
q=-(\ddot{a}/a)H^{-2}=-\left(1+\dot{H}H^{-2}\right)
=\frac{1}{2}+\frac{3}{2}\omega_\Lambda.
\end{equation}
\begin{figure}\center
\epsfig{file=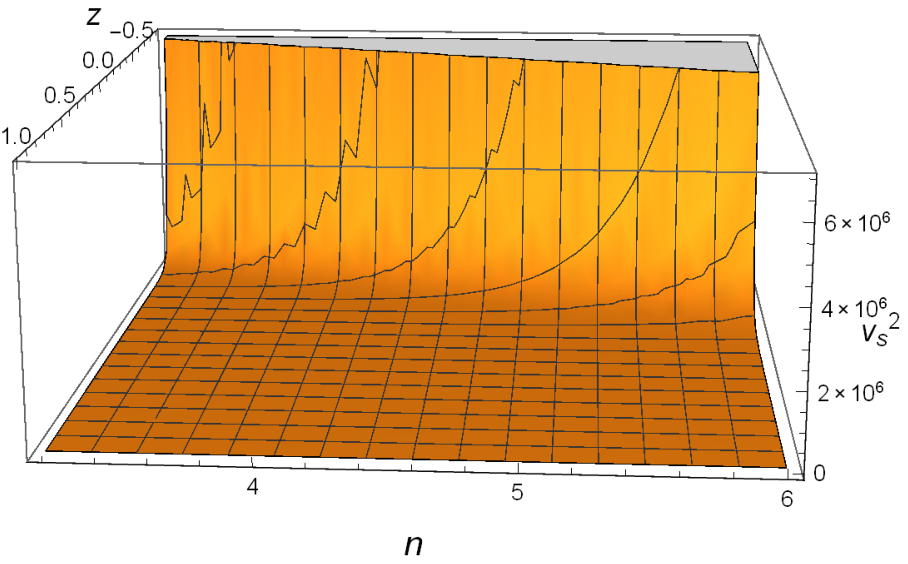,width=0.45\linewidth}
\epsfig{file=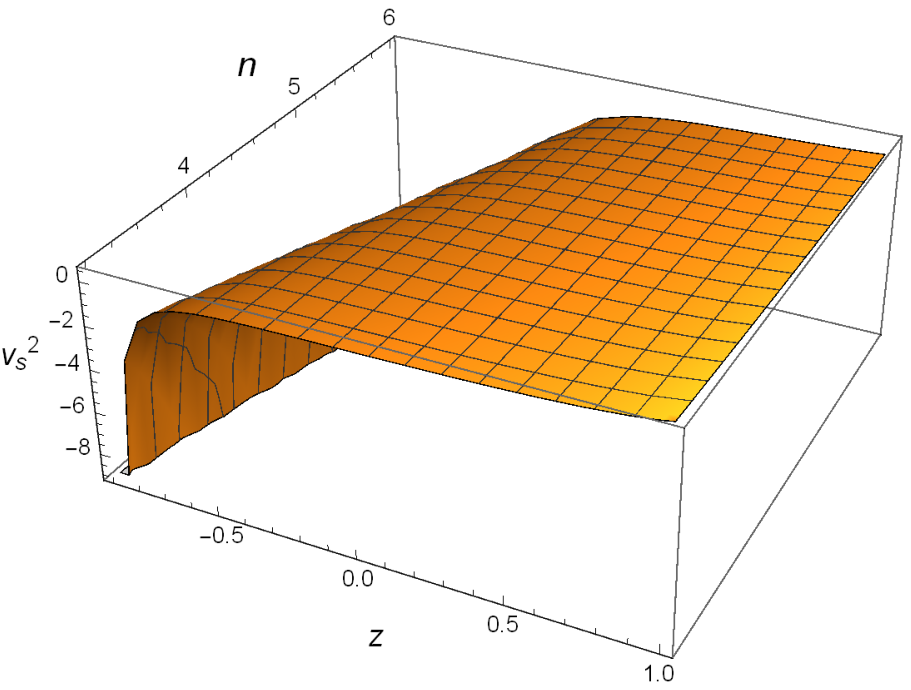,width=0.45\linewidth}\caption{Plots of power-law
$v_s^2$ for $u=2$ (left) and $-2$ (right).}
\end{figure}
\begin{figure}\center
\epsfig{file=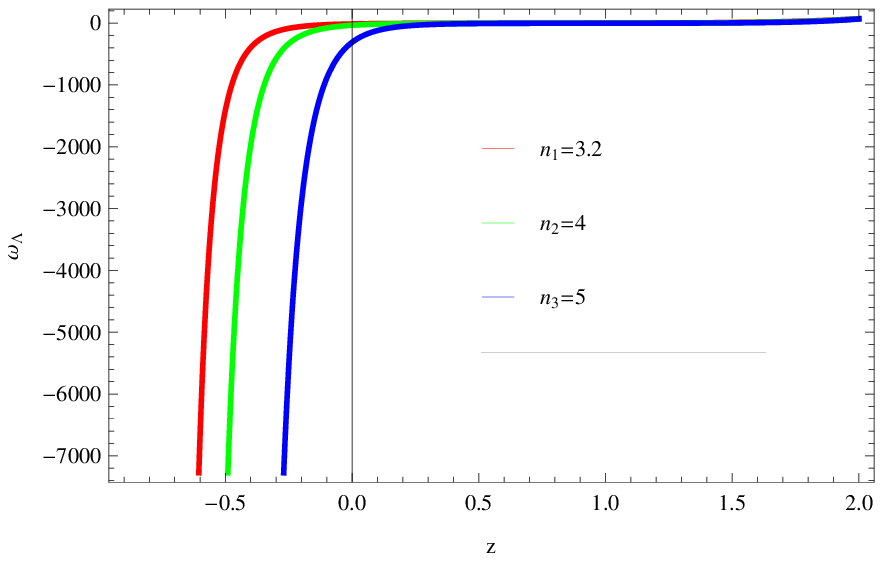,width=0.45\linewidth}
\epsfig{file=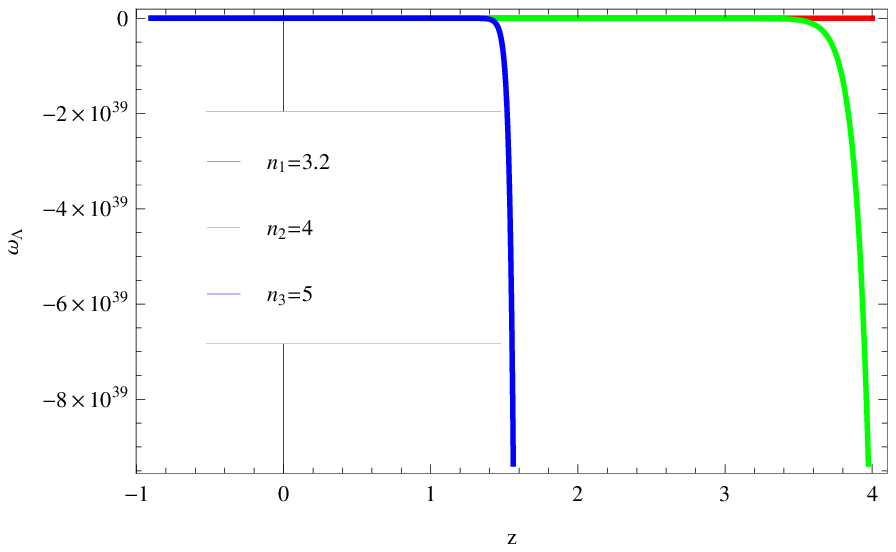,width=0.45\linewidth}\caption{Plots of power-law
EoS parameter $\omega_\Lambda$ for $u=2$ (left) and $u=-2$ (right).}
\end{figure}

Its positive value indicates decelerating behavior, $q=0$ expresses
constant expansion and negative value corresponds to accelerating
universe. Substituting the value of $\omega_\Lambda$ in the above
equation, we obtain deceleration parameter (Figure \textbf{4}). The
left plot for $u=2$ indicates that $q$ attains negative values in
the range $-0.9\leq z<0.15$, hence represents accelerating universe
in this interval. At $z=0.15$, it becomes zero showing constant
behavior and for $z>0.15$, it leads to decelerating universe. In the
right plot of Figure \textbf{4} $(u=-2)$, the deceleration parameter
exhibits negative values in the interval $-0.9\leq z\leq1.5$ showing
accelerating behavior.
\item The plane $\omega_{\Lambda}-\omega_{\Lambda'}$ is developed for
examining different DE models. Initially, Caldwell and Linder
\cite{11} used this method to study the behavior of quintessence DE
model. They suggested that the covered area in the phase plane
corresponds to two regions, thawing region
$(\omega_{\Lambda}<0,\omega_{\Lambda'}>0)$ as well as freezing
region $(\omega_{\Lambda}<0,\omega_{\Lambda'}<0)$. Here, we discuss
$\omega_{\Lambda}-\omega_{\Lambda'}$ plane corresponding to $u=2,-2$
for three different values of $n$. Figure \textbf{5} $(u=2)$ shows
that $\omega_{\Lambda}-\omega'_{\Lambda}$ plane represents thawing
regions for all three values of $n$. Similarly, all the curves
exhibit the same behavior for $u=-2$ in the interval $-0.9\leq
z\leq1$ as shown in Figure \textbf{6}. Hence,
$\omega_{\Lambda}-\omega'_{\Lambda}$ plane shows consistency with
the current behavior of the universe for $n=3.2,4$ and $5$.
\end{itemize}
\begin{figure}\center
\epsfig{file=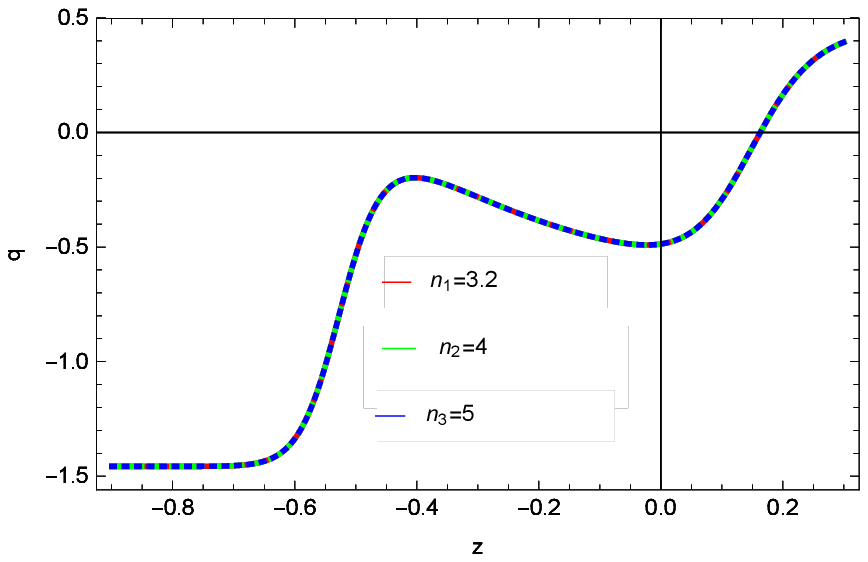,width=0.45\linewidth}
\epsfig{file=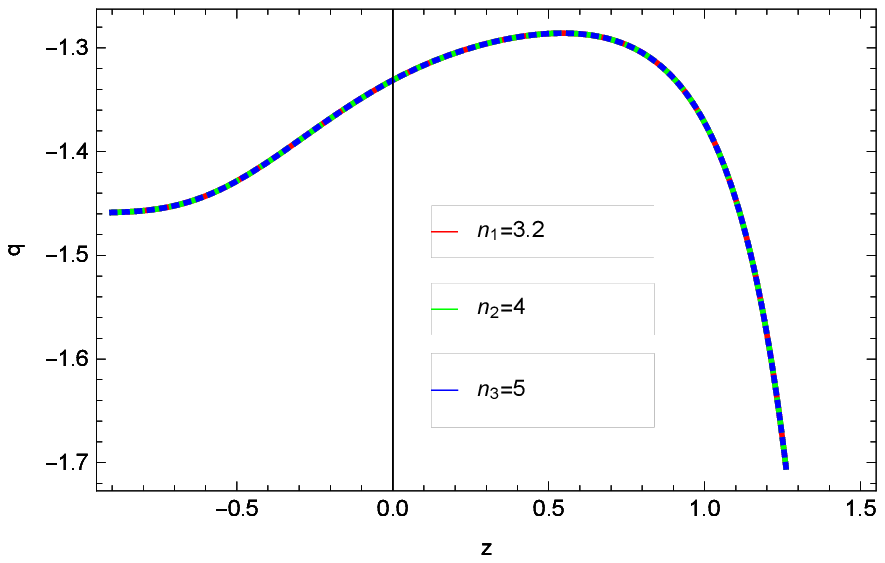,width=0.45\linewidth}\caption{Plots of power-law
deceleration parameter $q$ for $u=2$ (left) and $-2$ (right).}
\end{figure}
\begin{itemize}
\item Many DE models have been suggested to understand the
phenomenon of DE that ultimately explain the current behavior of the
universe. Some of them provide same values of the Hubble and
deceleration parameters. Thus it is necessary to determine which one
gives better information about acceleration of the expanding
universe. For this purpose, Sahni et al. \cite{20} presented two
dimensionless parameters named as statefinder parameters and are
defined as
\begin{figure}\center
\epsfig{file=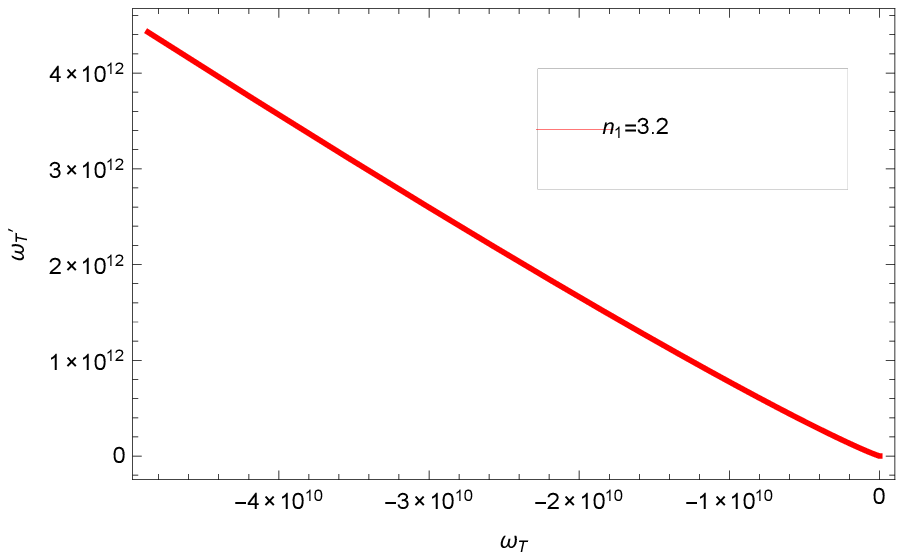,width=0.5\linewidth}\epsfig{file=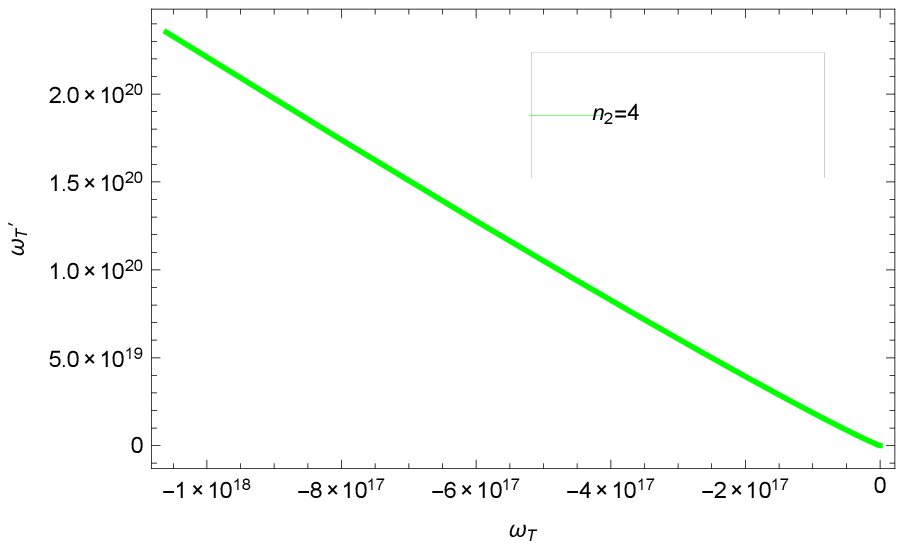,width=0.5\linewidth}
\epsfig{file=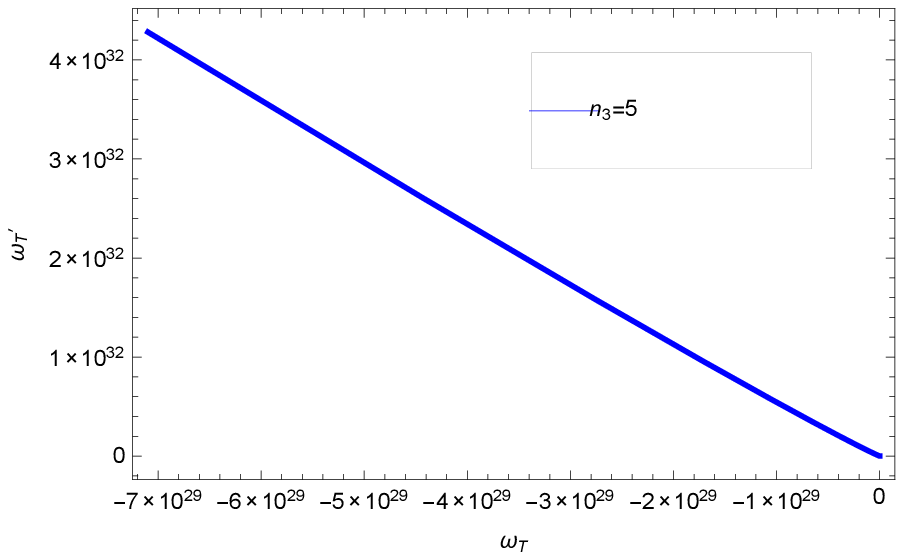,width=0.5\linewidth}\caption{Plots of power-law
$\omega_{\Lambda}-\omega_{\Lambda'}$ for $u=2$.}
\end{figure}
\begin{figure}\center
\epsfig{file=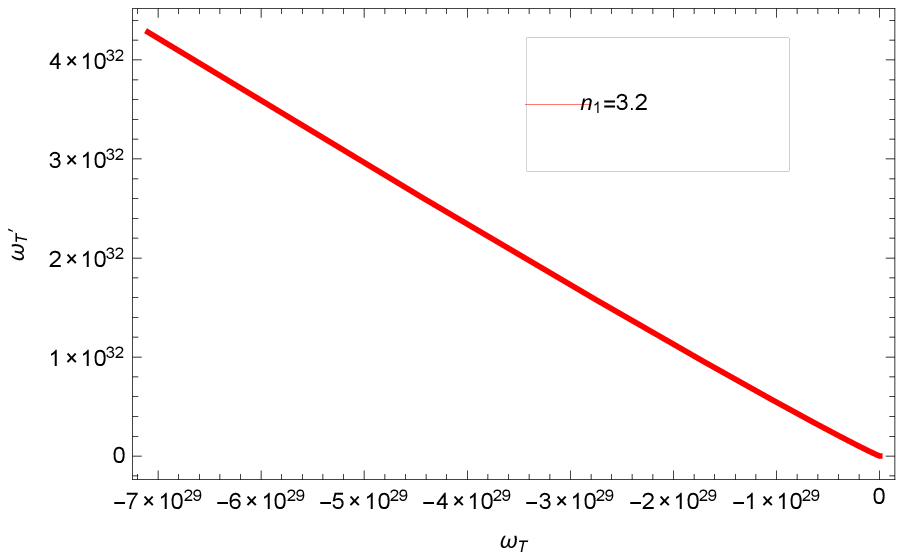,width=0.5\linewidth}\epsfig{file=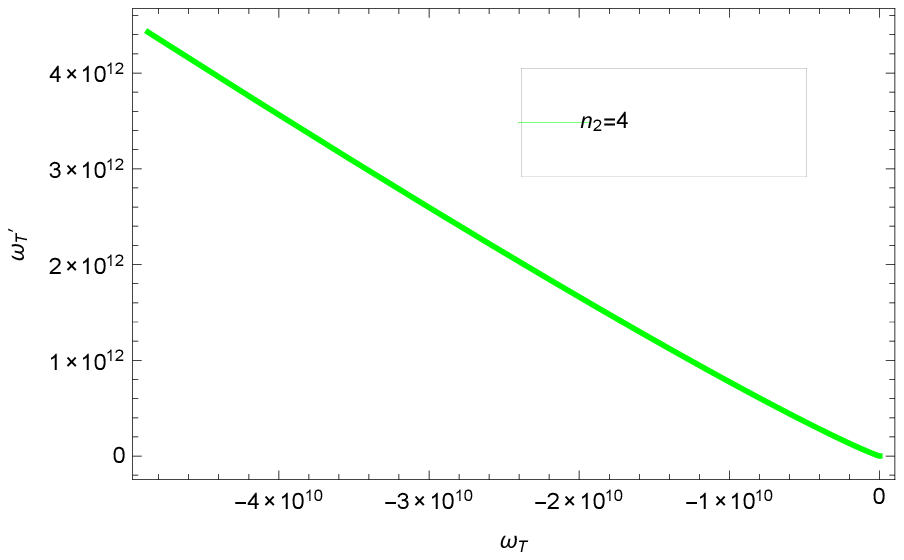,width=0.5\linewidth}
\epsfig{file=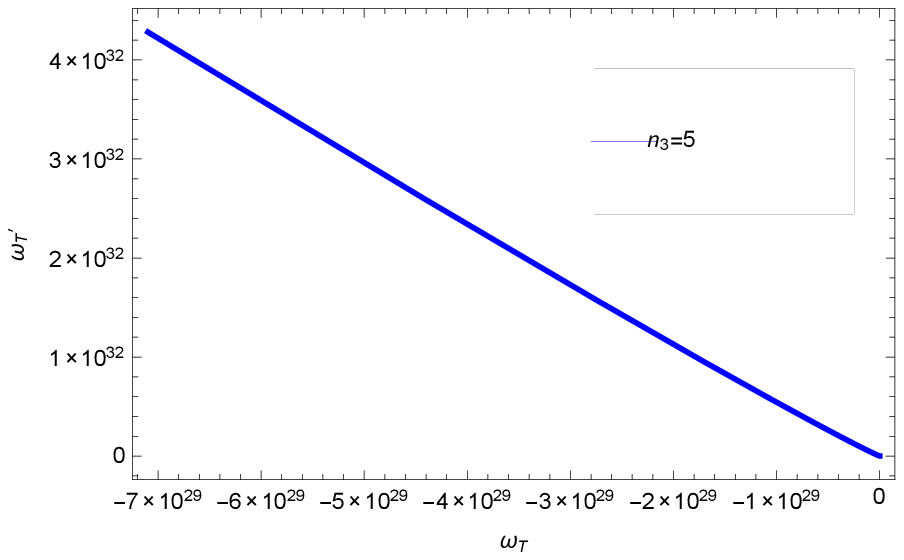,width=0.5\linewidth}\caption{Plots of power-law
$\omega_{\Lambda}-\omega'_{\Lambda}$ for $u=-2$.}
\end{figure}
\begin{equation}\nonumber
r=\frac{\dddot{a}}{a}\frac{1}{H^{3}},\quad
s=\frac{r-1}{(3q-\frac{3}{2})}.
\end{equation}
\begin{figure}\center
\epsfig{file=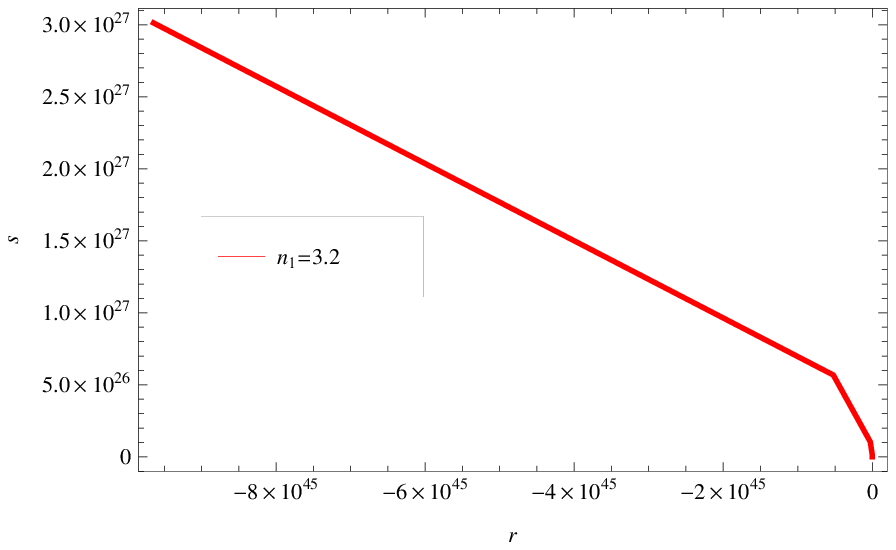,width=0.45\linewidth}\epsfig{file=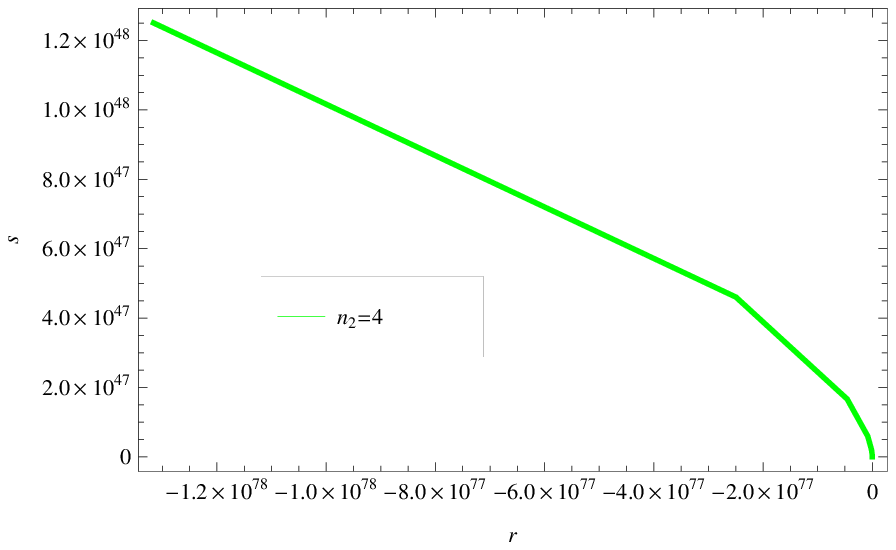,width=0.45\linewidth}
\epsfig{file=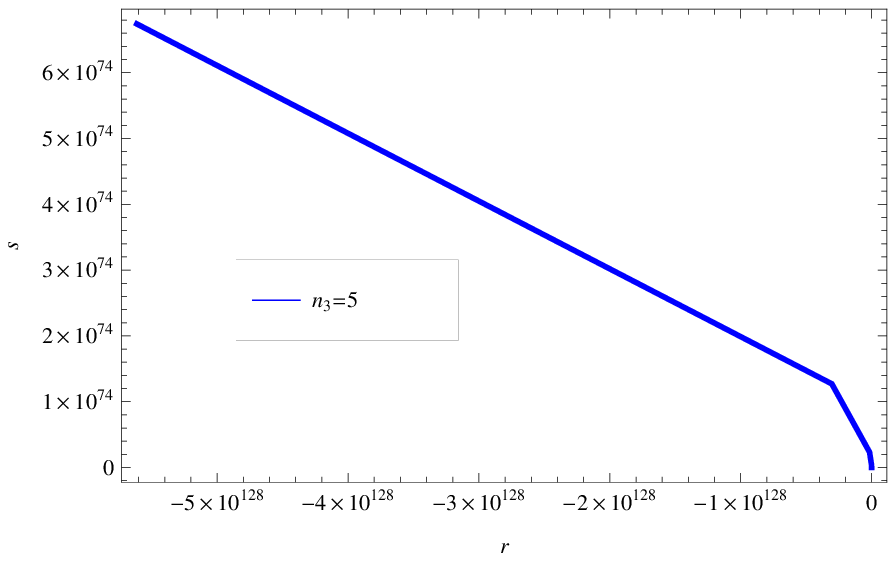,width=0.45\linewidth}\caption{Plots of power-law
$r-s$ plane for $u=2$.}
\end{figure}
We can also write the parameter $r$ in terms of $q$ as
$r=2q^2+q-q'$. These parameters describe the well-known cosmic
regions such as $(r,s)=(1,1)$ indicating CDM (cold dark matter)
limit and $(r,s)=(1,0)$ showing $\Lambda$CDM limit. The region
$s>0$, $r<1$ describes phantom and quintessence eras while $s<0$,
$r>1$ represents Chaplygin gas model. Here, we establish $r-s$
planes corresponding to our reconstructed GGPDE
$\tilde{F}(T,T_\mathcal{G})$ models for the same three values of
$n$. We assume the same values of corresponding parameters as in the
previous section in the range $-0.5\leq z\leq5$. Figures \textbf{7}
and \textbf{8} show that both reconstructed models for $u=2$ and
$-2$ provide the regions of quintessence and phantom DE eras as
$s>0$ and $r<1$ for all three values of $n$.
\end{itemize}

\subsection{Scale Factor for the Unified Phases}

Now, we investigate the behavior of $\tilde{F}(T,T_\mathcal{G})$
models through the scale factor for unified phases. For this
purpose, we substitute Eq.(\ref{12c}) in (\ref{12b}) with
$a=a_0(1+z)^{-1}$ to get a differential equation. We obtain the
reconstructed GGPDE $\mathcal{\tilde{F}}(T,T_\mathcal{G})$ model in
terms of $z$ by solving this differential equation numerically. For
both $u=2$ as well as $-2$, we consider $H_1=0.9$ and three
different values of $H_2=2.7,2.75,2.8$. Figure \textbf{9} shows that
both reconstructed models represent increasing behavior as the value
of $z$ increases.
\begin{figure}\center
\epsfig{file=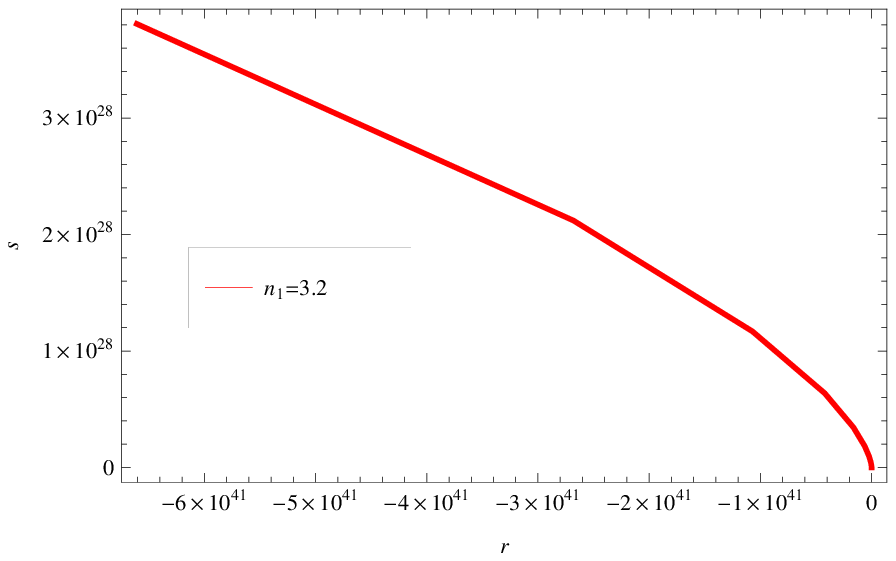,width=0.45\linewidth}\epsfig{file=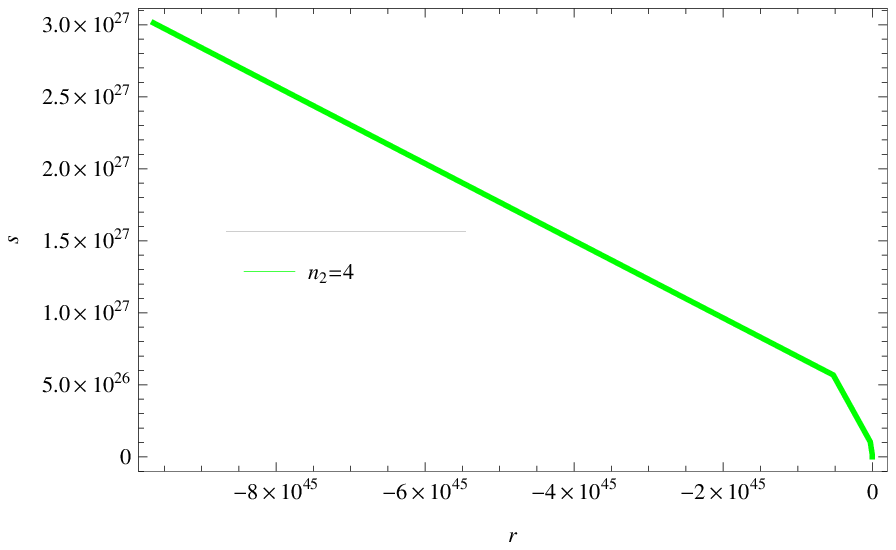,width=0.45\linewidth}
\epsfig{file=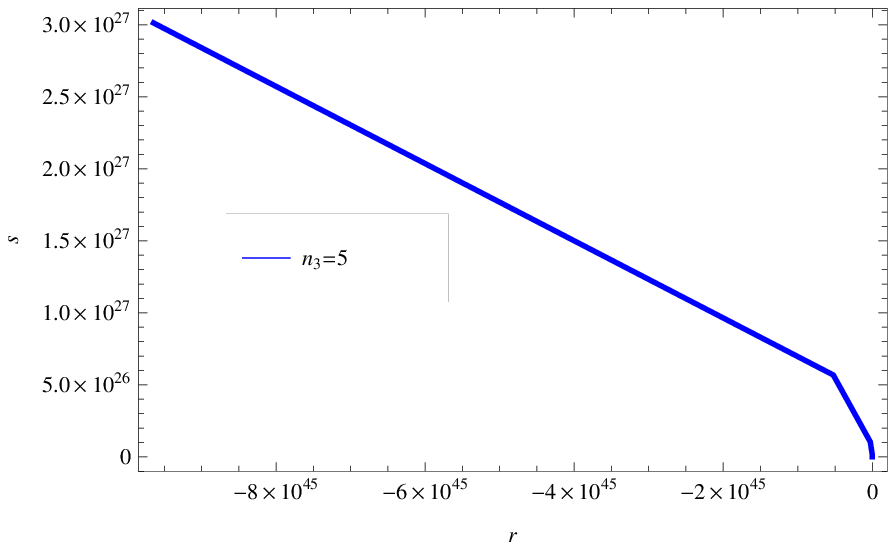,width=0.45\linewidth}\caption{Plots of power-law
$r-s$ plane for $u=-2$.}
\end{figure}
\begin{figure}\center
\epsfig{file=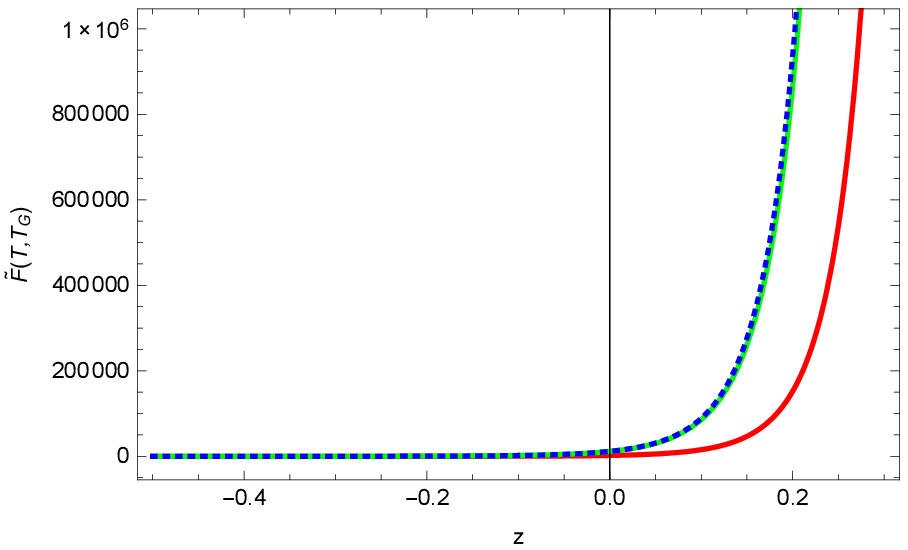,width=0.45\linewidth}
\epsfig{file=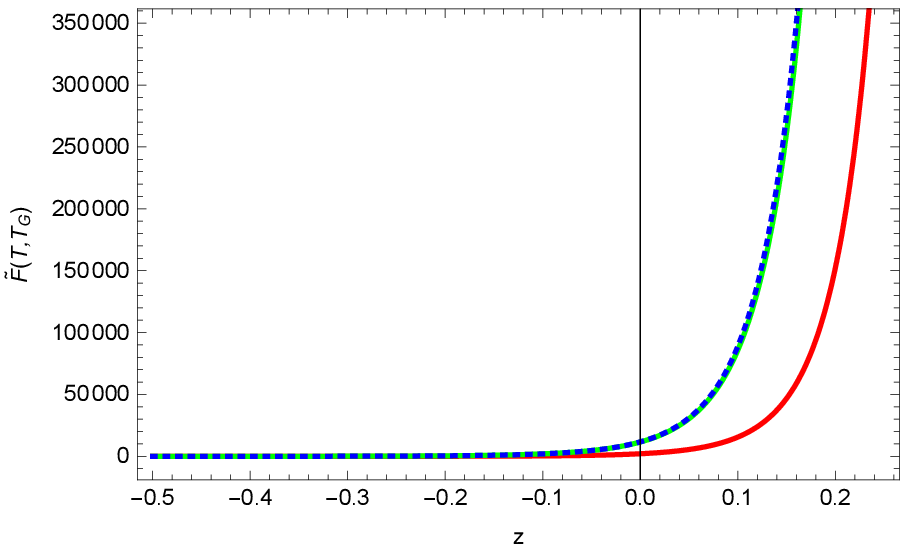,width=0.45\linewidth}\caption{Plots of unified
phases reconstructed GGPDE $\tilde{F}(T,T_\mathcal{G})$ model for
$u=2$ (left) and $-2$ (right). Also, $H_2=2.7$ (red), $H_2=2.75$
(green) and $H_2=2.8$ (blue).}
\end{figure}
\begin{figure}\center
\epsfig{file=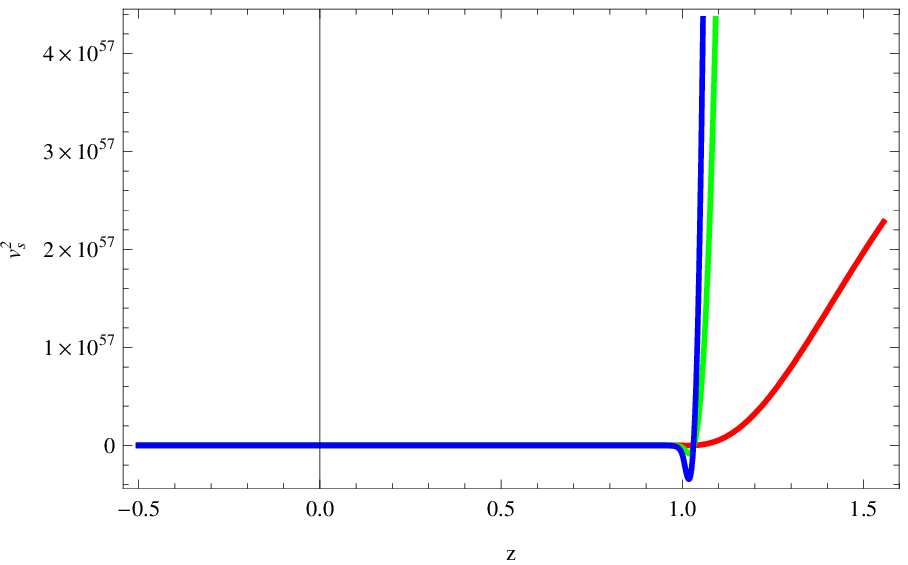,width=0.45\linewidth}
\epsfig{file=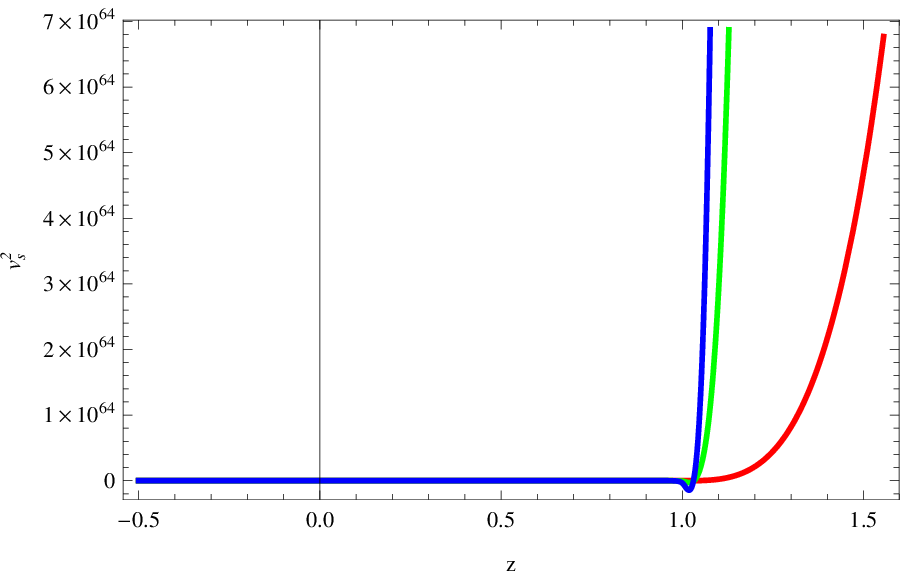,width=0.45\linewidth}\caption{Plots of unified
phases $v_s^2$ for $u=2$ (left) and $-2$ (right). Also, $H_2=2.7$
(red), $H_2=2.75$ (green) and $H_2=2.8$ (blue).}
\end{figure}
\begin{figure}\center
\epsfig{file=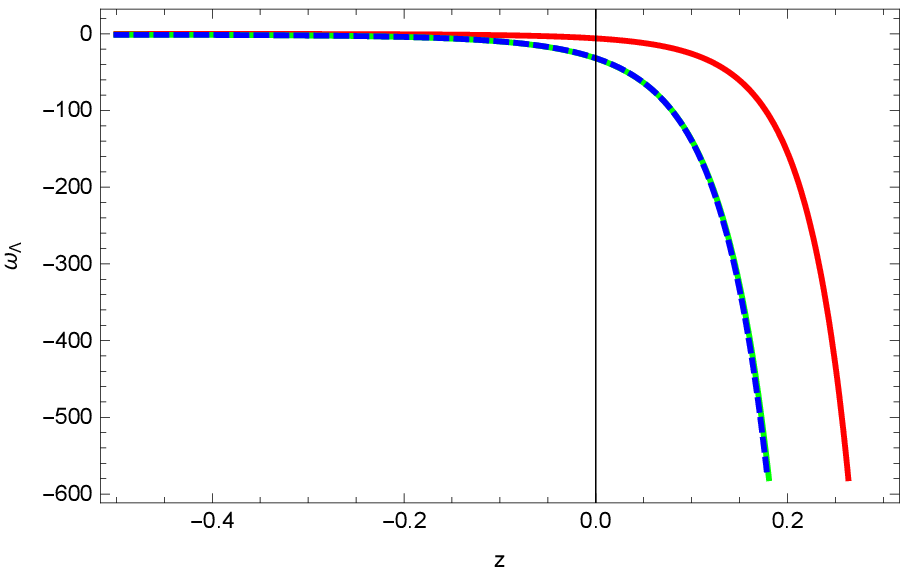,width=0.45\linewidth}
\epsfig{file=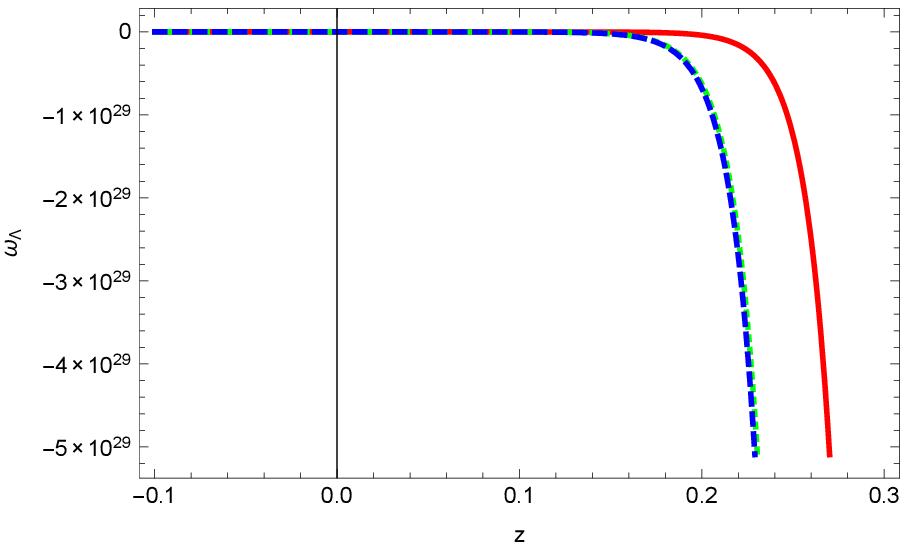,width=0.45\linewidth}\caption{Plots of unified
phases EoS parameter $\omega_\Lambda$ for $u=2$ (left) and $-2$
(right). Also, $H_2=2.7$ (red), $H_2=2.75$ (green) and $H_2=2.8$
(blue).}
\end{figure}
\begin{itemize}
\item Figure \textbf{10} represents the
behavior of squared speed of sound parameter for $u=2$ and $-2$.
Both plots show that positive values of $v_s^2$ for $z>1.02$ confirm
the stability of $\mathcal{\tilde{F}}(T,T_\mathcal{G})$ models.
\item Figure \textbf{11} indicates that EoS
parameter starts from matter dominated era initially
($\omega_\Lambda=0$) for $u=2$ as well as $-2$. At $z=0$,
$\omega_\Lambda$ crosses the phantom divide line for $u=2$ except
when $H_2=2.7$. For $u=-2$, the EoS parameter remains in the matter
dominated era for all values of $H_2$ and represents phantom
dominated era for $z>2.2$. However, in both cases, the EoS parameter
shows consistency with the current expanding behavior of the cosmos
as the value of $z$ increases.
\end{itemize}
\begin{itemize}
\item Figure \textbf{12} shows deceleration parameter in terms of
redshift parameter $z$. This is zero in the interval $-0.9\leq z
\leq1.1$ representing the constant cosmic expansion in both cases
$u=2$ as well as $-2$. However, for $z>1.1$, negative values of $q$
give rise to accelerating universe.
\item The behavior of $\omega_{\Lambda}-\omega'_{\Lambda}$ is shown
in Figure \textbf{13}. The left plot represents freezing region for
$H_2=2.7$ and $2.75$ while it corresponds to thawing region for
$H_2=2.8$. In the right plot, $\omega_{\Lambda}-\omega'_{\Lambda}$
expresses thawing region for $H_2=2.75$ and $2.8$.
\item The behavior of statefinder parameters is shown in Figure
\textbf{14}. In the left plot $(u=2)$, we notice that $s<0$, $r>1$
showing Chaplygin gas model while the right plot $(u=-2)$ indicates
$s>0$, $r<1$ implying phantom and quintessence eras of the universe.
\begin{figure}\center
\epsfig{file=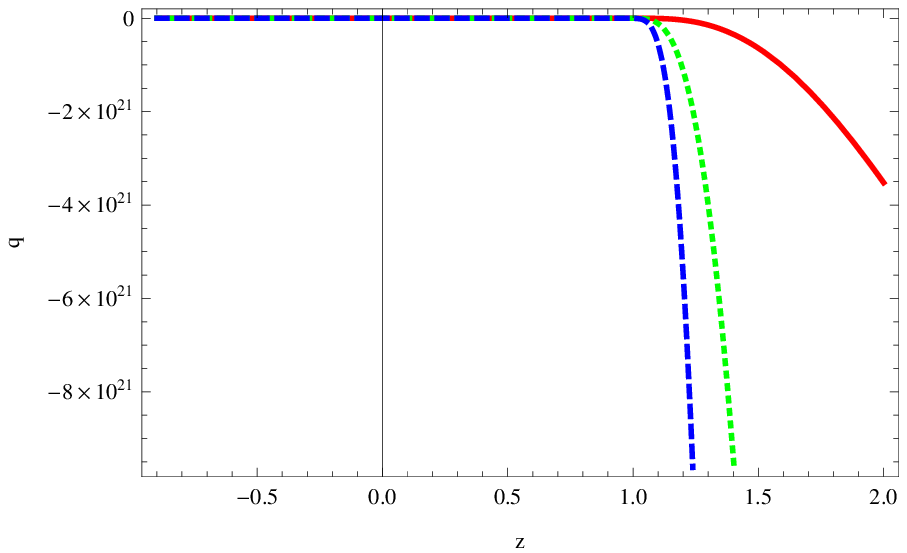,width=0.45\linewidth}
\epsfig{file=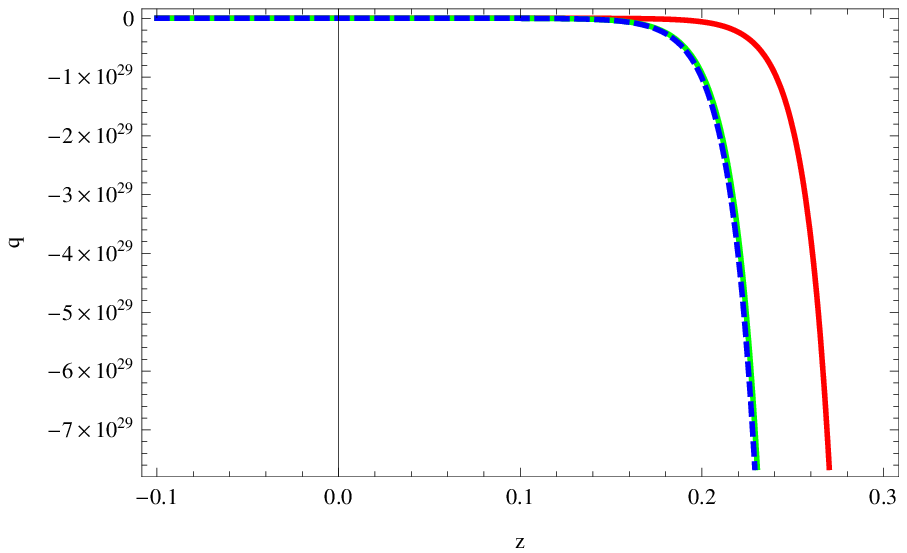,width=0.45\linewidth}\caption{Plots of unified
phases deceleration parameter $q$ for $u=2$ (left) and $-2$ (right).
Also, $H_2=2.7$ (red), $H_2=2.75$ (green) and $H_2=2.8$ (blue).}
\end{figure}
\begin{figure}
\center \epsfig{file=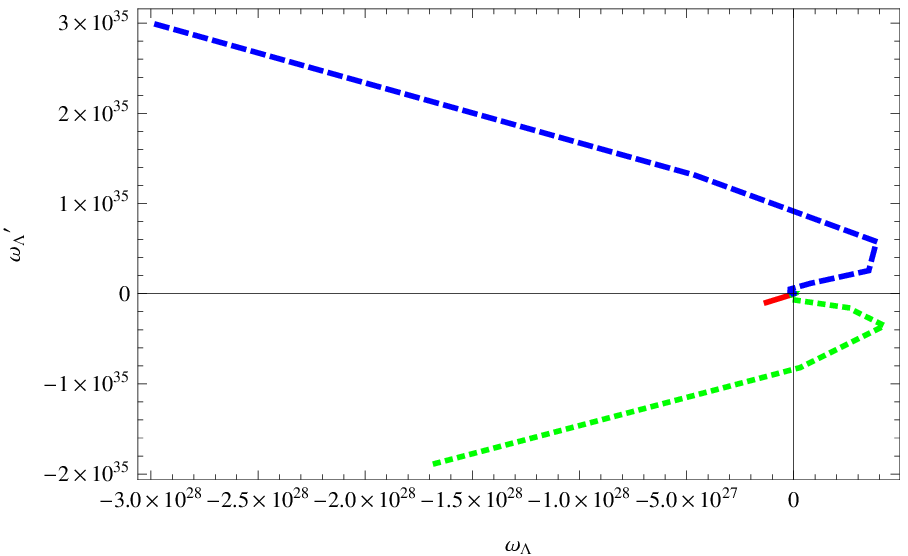,width=0.45\linewidth}
\epsfig{file=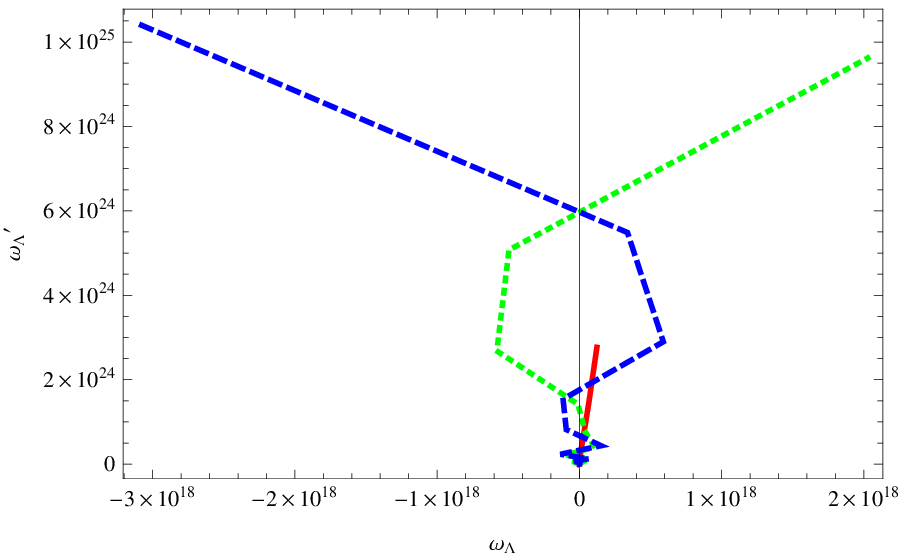,width=0.45\linewidth}\caption{Plots of unified
phases $\omega_{\Lambda}-\omega'_{\Lambda}$ for $u=2$ (left) and
$-2$ (right). Also, $H_2=2.7$ (red), $H_2=2.75$ (green) and
$H_2=2.8$ (blue).}
\end{figure}
\begin{figure}\center
\epsfig{file=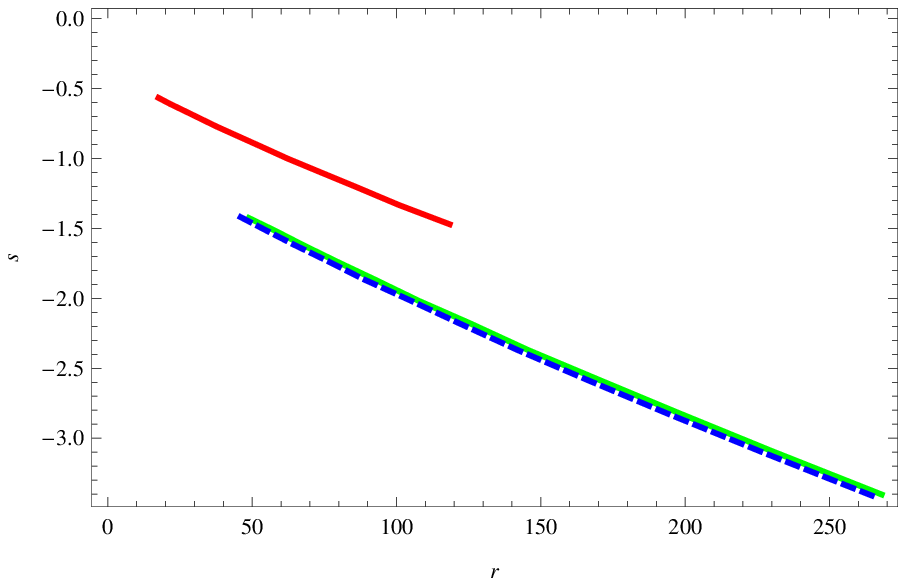,width=0.45\linewidth}
\epsfig{file=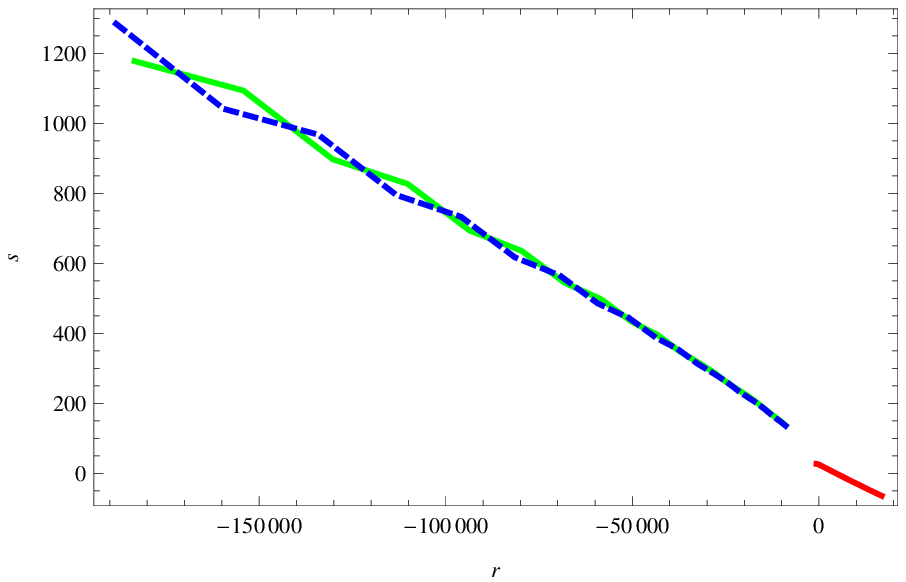,width=0.45\linewidth}\caption{Plots of unified
phases $r-s$ plane for $u=2$ (left) and $-2$ (right). Also,
$H_2=2.7$ (red), $H_2=2.75$ (green) and $H_2=2.8$ (blue).}
\end{figure}
\begin{figure}\center
\epsfig{file=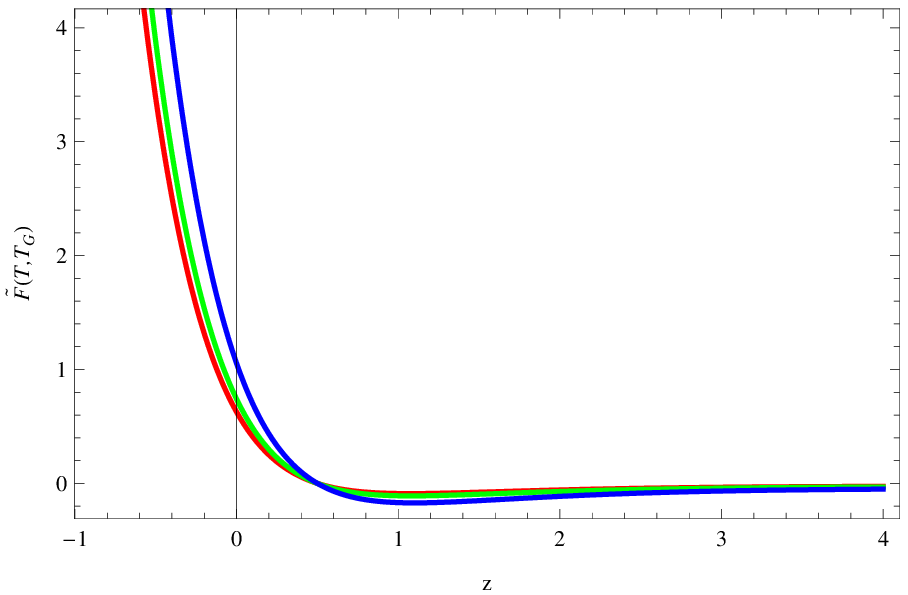,width=0.45\linewidth}
\epsfig{file=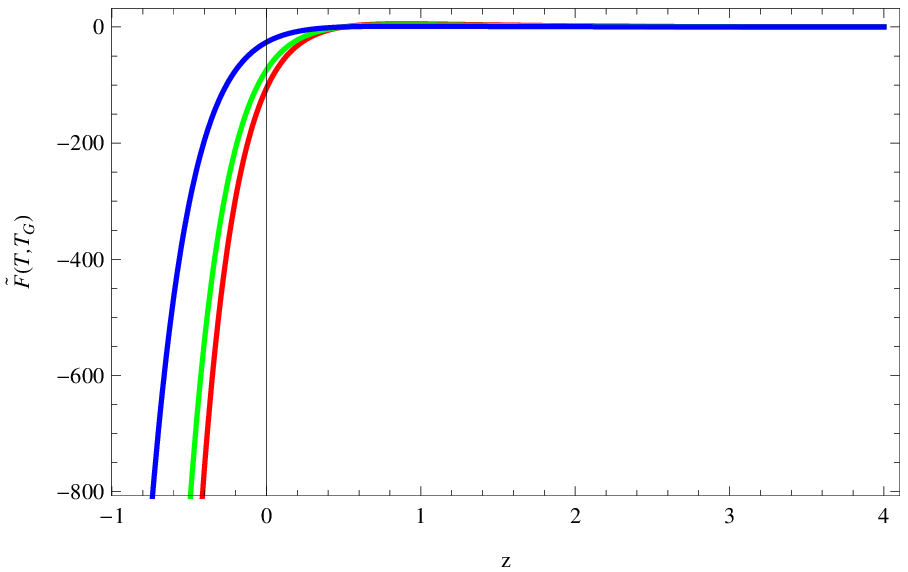,width=0.47\linewidth}\caption{Plots of
intermediate reconstructed GGPDE $\tilde{F}(T,T_\mathcal{G})$ model
for $u=2$ (left) and $-2$ (right).}
\end{figure}
\begin{figure}\center
\epsfig{file=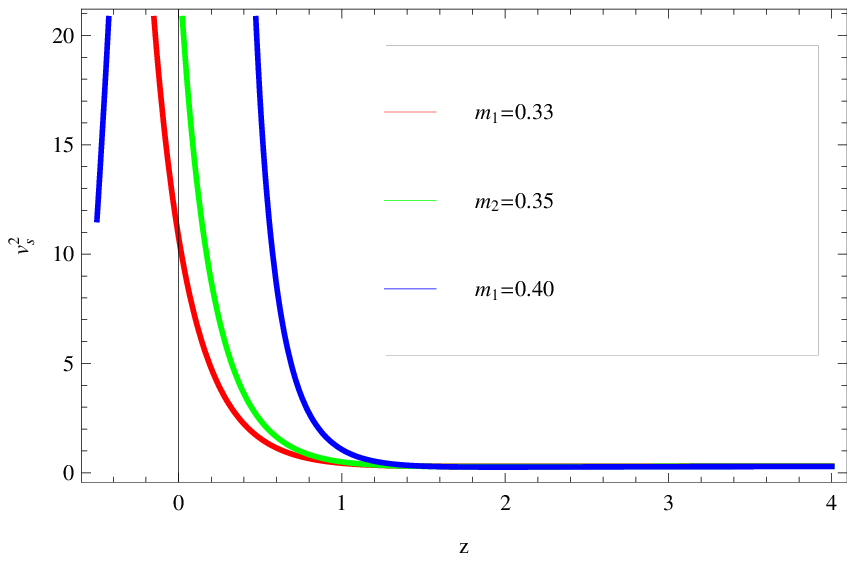,width=0.45\linewidth}
\epsfig{file=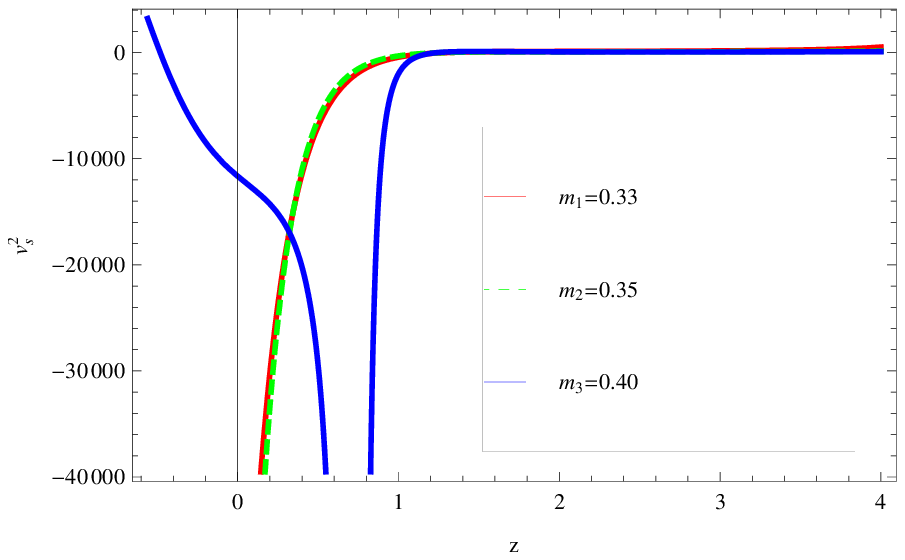,width=0.47\linewidth}\caption{Plots of
intermediate $v_s^2$ for $u=2$ (left) and $-2$ (right).}
\end{figure}
\begin{figure}\center
\epsfig{file=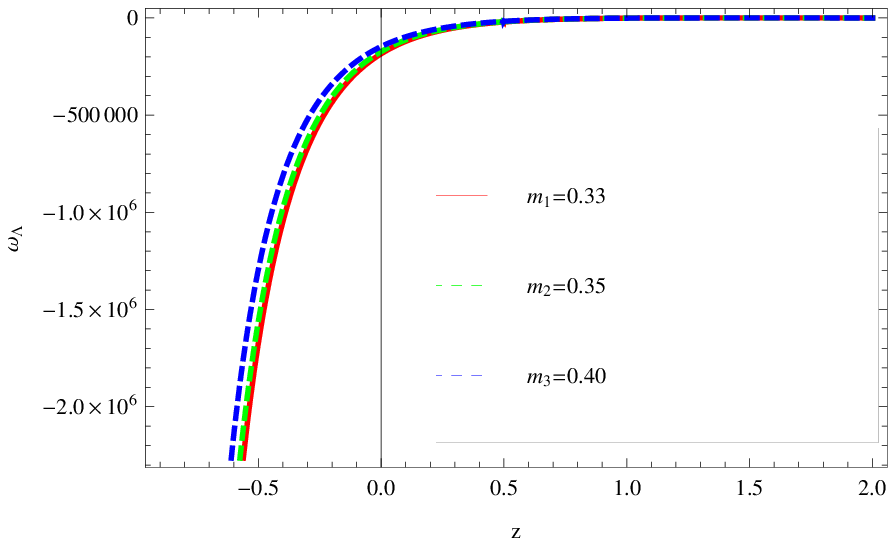,width=0.46\linewidth}
\epsfig{file=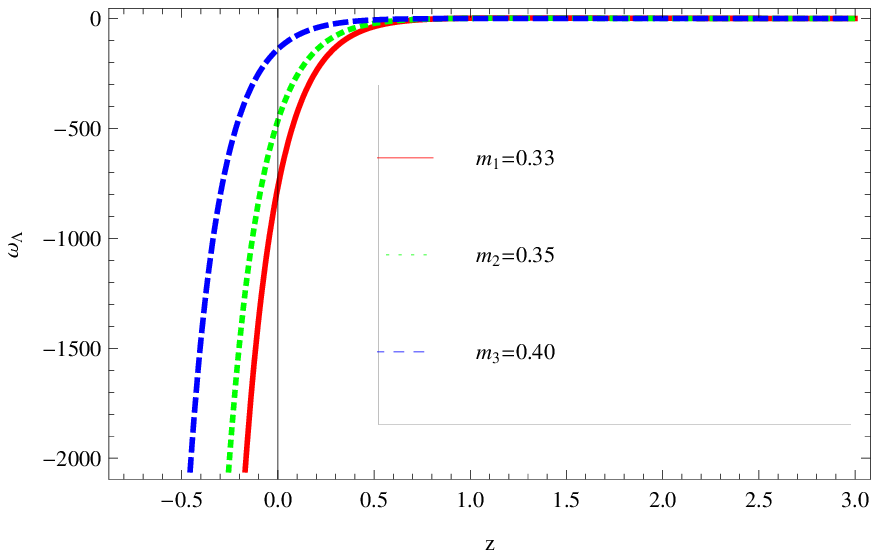,width=0.45\linewidth}\caption{Plots of
intermediate EoS parameter $\omega_\Lambda$ for $u=2$ (left) and
$-2$ (right). Also, $H_2=2.7$ (red), $H_2=2.75$ (green) and
$H_2=2.8$ (blue).}
\end{figure}
\begin{figure}\center
\epsfig{file=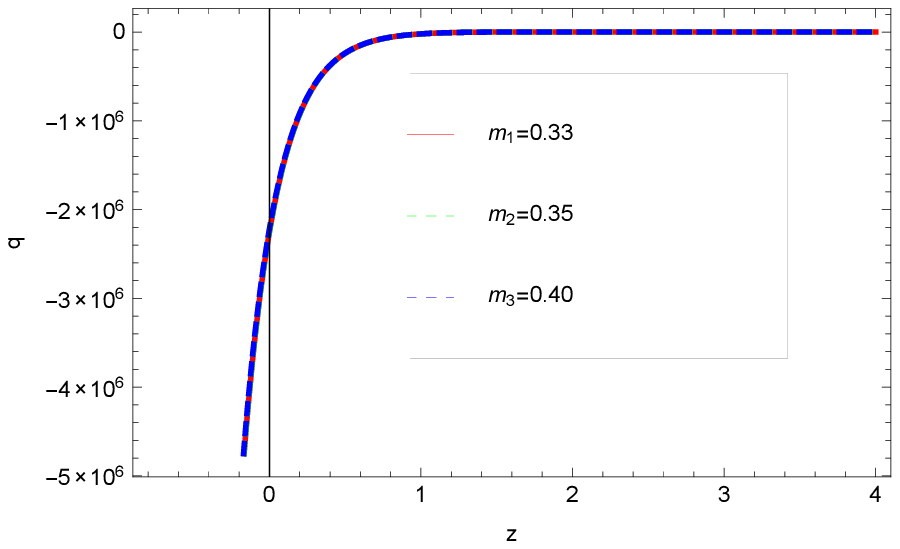,width=0.45\linewidth}
\epsfig{file=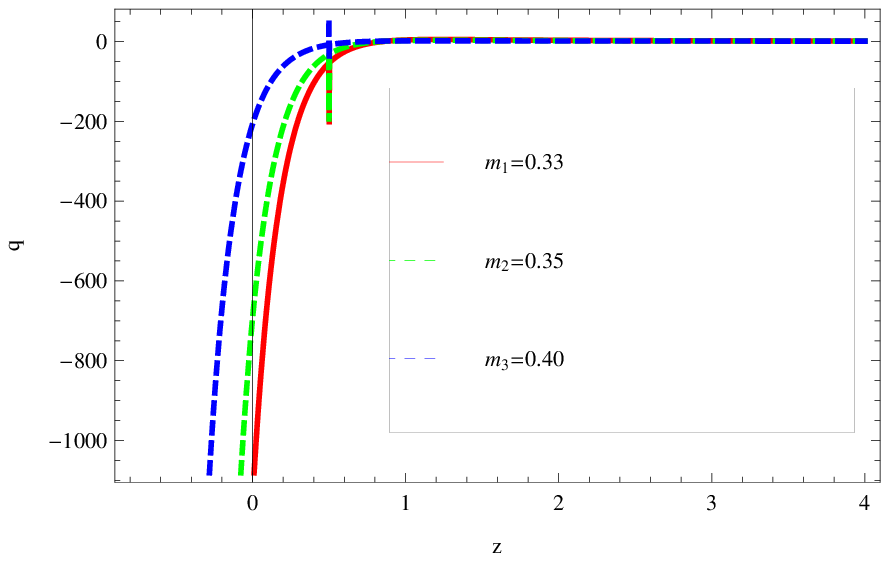,width=0.45\linewidth}\caption{Plots of
intermediate deceleration parameter $q$ for $u=2$ (left) and $-2$
(right). Also, $H_2=2.7$ (red), $H_2=2.75$ (green) and $H_2=2.8$
(blue).}
\end{figure}
\begin{figure}\center
\epsfig{file=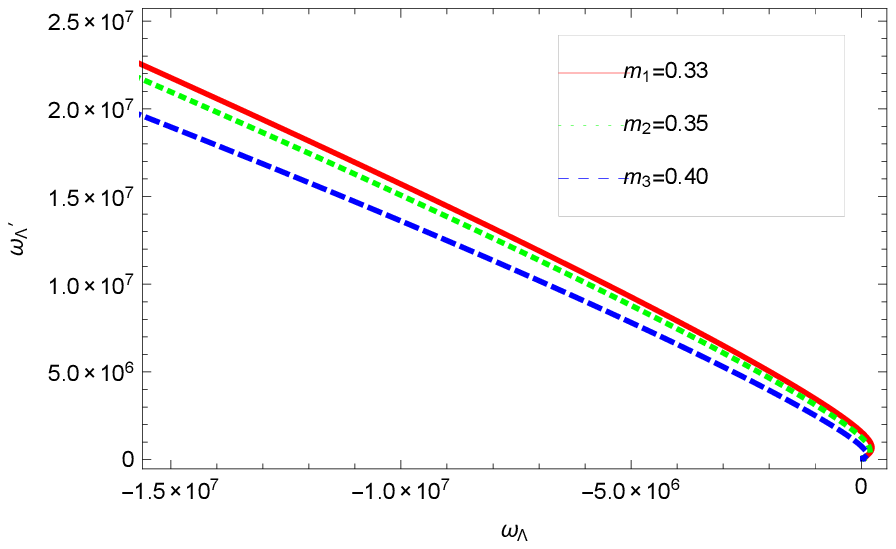,width=0.45\linewidth}
\epsfig{file=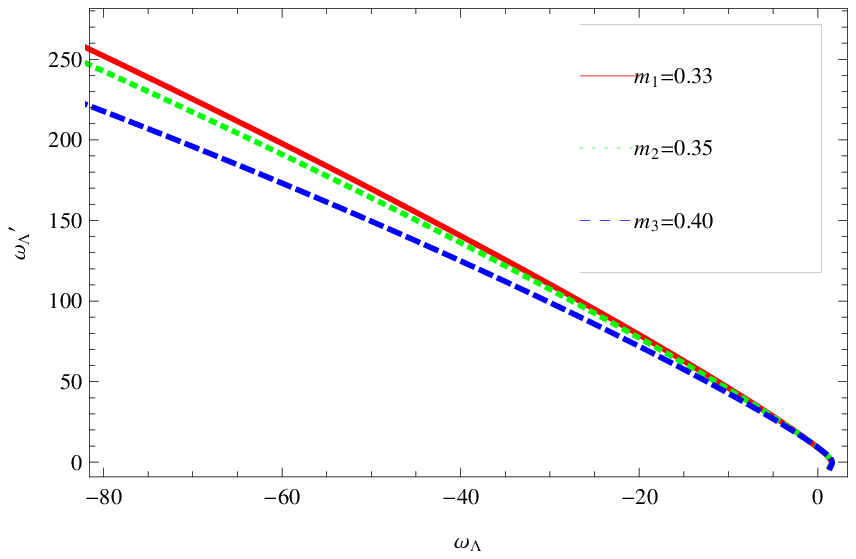,width=0.45\linewidth}\caption{Plots of
intermediate $\omega_{\Lambda}-\omega'_{\Lambda}$ for $u=2$ (left)
and $-2$ (right).}
\end{figure}
\begin{figure}\center
\epsfig{file=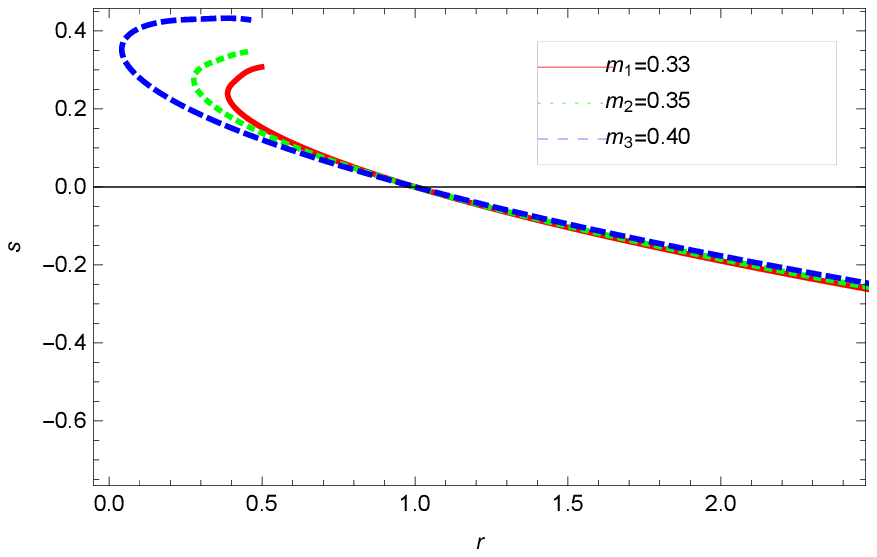,width=0.45\linewidth}
\epsfig{file=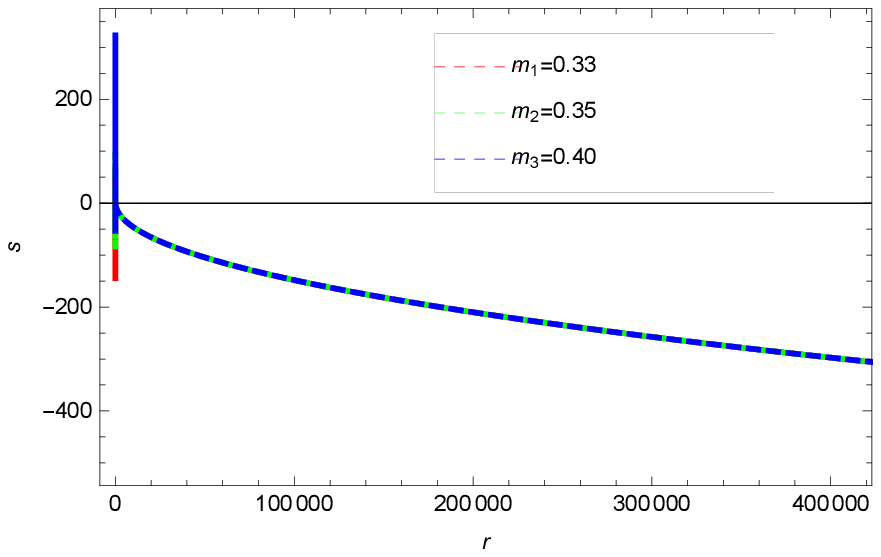,width=0.45\linewidth}\caption{Plots of
intermediate $r-s$ plane for $u=2$ (left) and $-2$ (right).}
\end{figure}
\end{itemize}

\subsection{Intermediate Scale Factor}

For this scale factor, we obtain reconstructed GGPDE
$\tilde{F}(T,T_\mathcal{G})$ models numerically by using
Eq.(\ref{12d}) in (\ref{12b}) as shown in Figure \textbf{15}. We
investigate the behavior of our models by assuming three different
values of $m$ as $m_1=0.33$, $m_2=0.35$, $m_3=0.40$ and choose
$b=0.1$. The left plot for $u=2$ represents decreasing behavior
while the right plot for $u=-2$ expresses increasing behavior.
\begin{itemize}
\item Figure \textbf{16} confirms the stability of our
corresponding model in the range $-0.9\leq z\leq1.15$ for $u=2$,
while for $u=-2$, the model shows instability.
\item Figure \textbf{17} indicates that
for both values of the PDE parameter, the EoS parameter exhibits
transition from phantom towards matter dominated era by crossing the
phantom divide line for all three values of $m$.
\item Figure \textbf{18} implies that both plots
attain negative values in the range $-0.9\leq z\leq1.8$ which
describe the accelerating behavior of the universe.
\item Figure \textbf{19} shows the plots for intermediate $\omega_{\Lambda}-\omega'_{\Lambda}$
plane that correspond to thawing regions.
\item In Figure \textbf{20} for $u=2$ (left), we attain the point
$(r,s)=(1,0)$ which shows $\Lambda$CDM limit. Also, the statefinder
parameters represent phantom and quintessence regions for $0\leq
z\leq1$. On the other hand, these parameters indicate Chaplygin gas
model for $1\leq z\leq2.5$. Similarly, the right plot for $u=-2$
shows Chaplygin gas model.
\end{itemize}

\section{Concluding Remarks}

In this paper, we have investigated the behavior of reconstruction
models \cite{10a} in $\mathcal{F}(T,T_\mathcal{G})$ gravity along
with GGPDE model as well as three scale factors. For this purpose,
we have considered two values of PDE parameter, i.e., $u=2$ and
$-2$. We have explored the role of some cosmological parameters
versus redshift parameter $z$ in this scenario. We have observed
that our models show stability for all the scale factors except
power-law and intermediate for $u=-2$. The equation of state
parameter in both cases $(u=2,-2)$ represents quintom behavior for
all the scale factors. It is found that reconstructed GGPDE
$\tilde{F}(T,T_\mathcal{G})$ models fulfill the condition of PDE
phenomenon. The plot of deceleration parameter versus $z$ exhibits
accelerated expansion of the universe.

We have observed that $\omega_{\Lambda}-\omega'_{\Lambda}$ plane
displays thawing region for power-law as well as intermediate scale
factor. The scale factor for two unified phases provides freezing
region for $H_2=2.8$. The $r-s$ plane shows phantom and quintessence
regions for power-law model $(u=2)$ as well as for the scale factor
of unified phases $(u=-2)$. For intermediate case, we have achieved
$\Lambda$CDM limit. We have found that both the planes, i.e.,
$\omega_{\Lambda}-\omega_{\Lambda'}$ as well as $r-s$ are consistent
with the current cosmic behavior.

Sharif and Jawad \cite{5d} analyzed GGPDE model by investigating its
cosmological consequences in general relativity. Our results are
consistent with them for non-interacting case with $\mu=1.55$ and
$\nu=1.91$. Jawad and Rani \cite{8} investigated the reconstructed
models, their stability and EoS parameter in the modified
Horava-Lifshitz $f(\tilde{R})$ gravity. Our results are in great
agreement with their work. Sharif and Nazir \cite{8a} discussed the
same cosmological parameters for GGPDE model in $\textit{f(T)}$
gravity. We have also compared our results with \cite{8a} and found
that the EoS parameter and cosmological planes for $u=2$ represent
consistency with the same values of model parameters. We have
noticed that EoS parameter is also consistent with the observational
data \cite{26} given as
\begin{eqnarray}\nonumber
\omega_\Lambda&=&-1.13^{+0.24}_{-0.25}\quad\text{(Planck
+WP+BAO)},\\\nonumber
\omega_\Lambda&=&-1.09\pm0.17\quad\text{(Planck+WP+Union
2.1)},\\\nonumber
\omega_\Lambda&=&-1.13^{+0.13}_{-0.14}\quad\text{(Planck+WP+SNLS)}.\\\nonumber
\omega_\Lambda&=&-1.24^{+0.18}_{-0.19}\quad\text{(Planck
+WP+$H_0$)},
\end{eqnarray}
These constraints have been evaluated by imposing various
observational techniques at $95\%$ level of confidence.

\end{document}